\documentclass[prl,twocolumn,superscriptaddress,nofootinbib]{revtex4-2}
\usepackage{epsfig}
\usepackage{graphicx}
\usepackage[english]{babel}
\usepackage{hyphenat}
\usepackage{amsmath}
\usepackage{amssymb}
\usepackage{mathtools}
\usepackage{mathrsfs}
\usepackage{bbm}
\usepackage{bm}
\usepackage{slashed}
\usepackage{epstopdf}
\usepackage{booktabs}
\usepackage{float}
\usepackage{placeins}
\usepackage{xfp}
\usepackage[table]{xcolor}
\definecolor{lcolor}{rgb}{0.,0.0,0.}
\definecolor{citcolor}{rgb}{0,0.,0.5}
\definecolor{darkgreen}{rgb}{0.0,0.5,0.0}
\usepackage[breaklinks,colorlinks,urlcolor=blue,citecolor=blue,linkcolor=blue]{hyperref}
\usepackage{multirow}
\usepackage{ltablex}
\usepackage{soul}
\usepackage{makecell}
\usepackage{ulem}
\usepackage{cleveref}

\def\be{\begin{eqnarray*}}
\def\ee{\end{eqnarray*}}
\def\beq{\begin{eqnarray}}
\def\eeq{\end{eqnarray}}

\newcommand{\bea}{\beq \begin{aligned}}
\newcommand{\eea}{\end{aligned}\eeq}

\newcommand{\Ncal}{\mathcal{N}}

\newcommand{\vect}[1]{\boldsymbol{#1}_{\perp}}

\newcommand{\bxt}{\vect{\bar x}}
\newcommand{\byt}{\vect{\bar y}}
\newcommand{\bzt}{\vect{\bar z}}
\newcommand{\brt}{\vect{\bar r}}

\newcommand{\pt}{p_T}
\newcommand{\qso}{Q_{s0}^2}
\newcommand{\gam}{\gamma}
\newcommand{\csq}{C^2}

\newcommand{\xo}{x_0}
\newcommand{\chisq}{\chi^2}

\def\be{\begin{eqnarray*}}
\def\ee{\end{eqnarray*}}
\def\beq{\begin{eqnarray}}
\def\eeq{\end{eqnarray}}

\newsavebox{\tabboxone}

\begin{document}

\title{Simultaneous Color Glass Condensate fit to deep inelastic scattering and forward hadron production at HERA, RHIC, and the LHC} 

\author{Piotr Korcyl}
\email{piotr.korcyl@uj.edu.pl}
\affiliation{Institute of Theoretical Physics, Jagiellonian University\\
prof. {\L}ojasiewicza 11, 30-348 Krak{\'o}w, Poland}
\author{Truong My Hau Le}
\email{truongmyhau.le@doctoral.uj.edu.pl}
\affiliation{Doctoral School of Exact and Natural Sciences, Jagiellonian University\\
prof. {\L}ojasiewicza 11, 30-348 Krak\'ow, Poland}
\author{Farid Salazar}
\email{farid.salazar@temple.edu}
\affiliation{Department of Physics, Temple University, Philadelphia, Pennsylvania 19122, USA}
\affiliation{RIKEN-BNL Research Center, Brookhaven National Laboratory, Upton, New York 11973, USA}
\affiliation{Physics Department, Brookhaven National Laboratory, Upton, New York 11973, USA}
\author{Tomasz Stebel}
\email{tomasz.stebel@uj.edu.pl}
\affiliation{Institute of Theoretical Physics, Jagiellonian University\\
prof. {\L}ojasiewicza 11, 30-348 Krak{\'o}w, Poland}

\begin{abstract}

We present the first simultaneous fit to deep inelastic scattering (DIS)
reduced cross-sections from HERA and forward single inclusive hadron
production (SIHP) from RHIC and the LHC in which the dipole amplitude is
obtained from Balitsky--Kovchegov (BK) evolution within the Color Glass
Condensate effective theory. We demonstrate that, using the LO BK equation
with running coupling (with or without kinematical constraint) and a
constant per-collider $K$-factor accounting for higher-order corrections, a
global $\chi^2/\text{d.o.f.}$ close to unity is achieved in both schemes. The required $K$-factors are
about a factor of two larger at RHIC than at the LHC, in
agreement with threshold-resummed one-loop studies of forward production.
In our setup and with the data available, we find no tension between the
two observables: the amplitude determined by DIS alone reproduces the
$\pt$ and rapidity shapes of the forward hadron spectra from RHIC to LHC
energies, and the global-fit parameters remain unchanged within
uncertainties. We further quantify the
dependence of the $K$-factors on the fragmentation-function set and the
factorization scale, identifying these as the dominant systematic
uncertainties to be addressed in future precision analyses of forward
hadron production. We perform a full uncertainty analysis using the Hessian
method, validated against a Monte Carlo Bayesian analysis.

\end{abstract}


\maketitle
\textbf{Introduction.} Quantum Chromodynamics (QCD) is the fundamental theory of strong interactions. A major goal is mapping the properties of hadrons, strongly interacting bound states, in terms of the basic constituents of QCD: quarks and gluons~\cite{Feynman:1969ej,Bjorken:1969ja}. Deep
Inelastic Scattering (DIS) experiments at HERA have elucidated the partonic picture; for example, by making it possible to extract parton distribution functions of the proton. These experiments revealed that gluons are copious at high energies, carrying a small fraction of their parent hadron's momentum. At sufficiently high energies, it is expected that the linear evolution equations, DGLAP \cite{Gribov:1972ri,Altarelli:1977zs,Dokshitzer:1977sg} and BFKL \cite{Lipatov:1976zz,Kuraev:1977fs,Balitsky:1978ic}, must be amended by the introduction of non-linear
recombination effects~\cite{Gribov:1984tu,Mueller:1985wy,McLerran:1993ni},
leading to a phenomenon known as gluon saturation and the emergence of a
dynamically generated scale, the saturation scale $Q_s(x)$. There have been many
hints of saturation in different observables across various collider
experiments, but its unambiguous identification remains elusive due to competing
QCD effects that mask saturation effects~\cite{Morreale:2021pnn}. The clear
observation of gluon saturation is one of the major experimental goals of
facilities such as the Relativistic Heavy Ion Collider and the Large Hadron
Collider, and is a central objective of the future Electron-Ion
Collider~\cite{AbdulKhalek:2021gbh}. The Color Glass Condensate
(CGC)~\cite{Gelis:2010nm} provides an effective theory for this regime of QCD.
Partonic cross-sections in the CGC are expressed as convolutions of Wilson-line
correlators with perturbatively calculable impact factors. While the initial
condition of these correlators is non-perturbative, their high-energy/rapidity
evolution is perturbatively calculable and dictated by non-linear
renormalization group equations, notably the Balitsky--Kovchegov (BK) equation and
the Jalilian-Marian--Iancu--McLerran--Weigert--Leonidov--Kovner (JIMWLK)
equation~\cite{Balitsky:1995ub,Kovchegov:1999yj,JalilianMarian:1997jx,JalilianMarian:1997gr,Kovner:2000pt,Iancu:2000hn,Iancu:2001ad,Ferreiro:2001qy}.

An attractive feature of the CGC is that it can describe heterogeneous observables across colliding systems — electron-nucleus, proton-nucleus, and nucleus-nucleus — from the same building blocks~\cite{Albacete:2014fwa,Morreale:2021pnn}.
Its simplest manifestation is that both the reduced
cross-section for DIS and the forward single
inclusive hadron production (SIHP) in proton-proton collisions depend on the
dipole amplitude, the simplest correlator of Wilson
lines~\cite{Kovchegov:1999yj,Dumitru:2005gt,Gelis:2002nn}. Predictions for both observables exist at next-to-leading order (NLO) accuracy
within the CGC effective theory~\cite{Beuf:2020dxl,Chirilli:2012jd,Liu:2019iml,Hanninen:2022gje,Casuga:2026xxt,Shi:2021hwx,Mantysaari:2023vfh}.

Traditionally, the non-perturbative parameters in the initial conditions are
constrained by fitting the HERA reduced DIS cross-section
\cite{Albacete:2009fh,Albacete:2010sy,Rezaeian:2013tka,Beuf:2020dxl,Casuga:2026xxt}, after which predictions are made
for other observables \cite{Goncalves:2006yt,Albacete:2010bs,Lappi:2013zma,Kutak:2012rf,Salazar:2021mpv,
Benic:2022ixp,Caucal:2025zkl,Fujii:2026ccu}. There have also been efforts
describing several observables simultaneously within DIS
\cite{Kowalski:2006hc,Rezaeian:2013tka,Bendova:2019psy,Hanninen:2022gje,Dai:2026nzp,Kou:2026iau}
or ultra-peripheral collisions \cite{Mantysaari:2025ltq,Mantysaari:2026vgx}, but attempts at describing two distinct observables across very different colliding systems are missing. The goal of this work is to perform
the first simultaneous CGC-based fit of the reduced DIS cross-section from HERA
and SIHP in proton-proton collisions from RHIC and LHCb to fill this gap. Our present analysis is
restricted to the leading-order (LO) impact factor for both observables, with
the running-coupling BK equation for the evolution of the dipole amplitude. Our
motivation for this joint fit is three-fold: (i) to test the universality of the
dipole amplitude across these two channels at the level of its shape and
evolution; (ii) to identify potential tensions and assess the constraining power
in the description of these two distinct observables; and (iii) to estimate the
role of higher-order corrections and the uncertainty due to non-perturbative
inputs.

Since higher-order corrections can be sizable, we introduce
$K$-factors as a simple proxy to estimate these missing contributions. We
adopt this approach rather than a full NLO treatment because the fixed-order
NLO cross-section for SIHP is known to turn negative at large transverse
momentum \cite{Stasto:2013cha}, with results sensitive to the treatment of the
rapidity divergence and to the scale prescription
\cite{Stasto:2014sea,Altinoluk:2014eka,Watanabe:2015tja,Ducloue:2016shw,Iancu:2016vyg,Ducloue:2017mpb}; threshold resummation restores positivity and
enhances the cross-section \cite{Liu:2020mpy,Shi:2021hwx}. 
NLO corrections for SIHP are expected to be larger at RHIC than at the LHC, since RHIC's lower energy probes momentum fractions closer to the kinematic threshold, where threshold resummation produces correspondingly larger
corrections~\cite{Shi:2021hwx}. We therefore introduce separate $K$-factors for RHIC and the LHC, a reasonable simplification since this energy dependence should dominate over the $K$-factor's milder $\pt$ and rapidity dependence.

\textbf{The dipole, DIS and SIHP. }The basic building block in the CGC is the two-point correlator (dipole) of Wilson lines,
\begin{align}
    \mathcal{D}_{R}(r_T, x) = \frac{1}{N_c} \left \langle \mathrm{Tr}\left[ V_R(\boldsymbol{x}) V^\dagger_R(\boldsymbol{y}) \right] \right \rangle_x\,,
\end{align}
where $R$ is the $SU(3)$ representation and, assuming translational invariance, the coordinate dependence reduces to the separation $r = |\boldsymbol{x}-\boldsymbol{y}|$. The $x$-dependence of the two-point function is given by the Balitsky-Kovchegov equation \cite{Balitsky:1995ub,Kovchegov:1999yj}.

At LO in the CGC, the DIS cross-section can be obtained from the optical theorem by computing the forward scattering amplitude for a $\gamma^*$-proton collision:
\begin{align}
& \sigma_{T,L}^{\gamma^* p}(x_{\mathrm{Bj}},Q^2) \nonumber\\
&= \frac{\sigma_0}{2}  \int  \mathrm{d}^2 \boldsymbol{r} \ 2 \mathcal{N}_F(r, x_{\mathrm{Bj}})  \int \mathrm{d} z |\psi_{T,L}^{\gamma^* \rightarrow f \bar{f}}(z,r,Q^2)|^2 \,.
\end{align}
Here $\psi_{T,L}^{\gamma^* \rightarrow f \bar{f}}(z,r,Q^2)$ is the perturbatively calculable light-cone wave function for a photon with virtuality $Q^2$ fluctuating into a $f\bar{f}$ dipole of transverse size $r$, with the quark carrying longitudinal momentum fraction $z$ of the virtual photon.

The underlying QCD dynamics is captured by the dipole amplitude $\mathcal{N}_F(r, x_{\mathrm{Bj}})=1-\mathcal{D}_F(r, x_{\mathrm{Bj}})$ in the fundamental representation, which describes the scattering of the quark-antiquark pair with the small-$x$ dense gluon field of the proton. The Bjorken-$x$ variable is $x_{\mathrm{Bj}} \approx Q^2/W^2$, with $W$ the photon-proton center-of-mass energy. Neglecting the dipole amplitude's impact-parameter dependence, the $\boldsymbol{b}$-integration is replaced by the effective proton transverse area $\sigma_0/2$.

The reduced cross-section is given by
\begin{align}
    \sigma_{r} (y,x,Q^2) = F_2(x,Q^2) - \frac{y^2}{1+(1-y)^2} F_L(x,Q^2)\,,
\end{align}
where $y=W^2/s$ is the DIS inelasticity, and $s$ is the squared center-of-mass energy of the electron-proton system. The structure functions are defined as $F_2(x,Q^2) = \frac{Q^2}{4 \pi^2\alpha_{em}}(\sigma_T^{\gamma^* p}+\sigma_L^{\gamma^* p})$ and $ F_L(x,Q^2) = \frac{Q^2}{4 \pi^2\alpha_{em}}\sigma_L^{\gamma^* p}$, with $T$ and $L$ the transverse and longitudinal photon components.

Let us now turn to SIHP. In the CGC at
leading order, the differential cross-section for SIHP in $pp$ collisions can be
written in $k_T$-factorized form as a convolution of the unintegrated gluon
distributions of the two colliding protons~\cite{Blaizot:2004wu,Blaizot:2004wv}. When the produced hadron's transverse momentum is parametrically larger than one proton's saturation scale, that proton's transverse-momentum contribution can be neglected. This dilute-dense approximation~\cite{Dumitru:2005gt} suits forward particle production, where the two protons probe very different $x$ values: $x_1 =
\frac{p_T}{z\sqrt{s}}e^{y_h} \sim 1$ and $x_2 = \frac{p_T}{z\sqrt{s}}e^{-y_h}
\ll 1$ (these expressions hold for light hadrons, where the difference between the rapidity $y_{h}$ and pseudo-rapidity $\eta_{h}$ is negligible). One proton is therefore dilute (small saturation scale) and the other
dense (large saturation scale). For a recent quantitative comparison between
$k_T$-factorization and the dilute-dense approximation,
see~\cite{Fujii:2026ccu}. In the dilute-dense approximation, the large-$x$
partons in the proton are treated as collinear and are described by parton
distribution functions (PDFs) $f_{i/p}$. The scattering of these partons with the low-$x$ gluon field is encoded in the Fourier transform of the dipole, $\widetilde{\mathcal{D}}_{R}(\boldsymbol{k},x) = \int \mathrm{d}^2 \boldsymbol{r}\ e^{-i \boldsymbol{k} \cdot \boldsymbol{r}} \mathcal{D}_{R}(\boldsymbol{r}, x)$.
The scattered parton then fragments via collinear fragmentation functions (FFs) $D_{h/i}$. The differential cross-section at leading order in the dilute-dense framework reads~\cite{Dumitru:2005gt,Chirilli:2012jd}:
\begin{align}
& \frac{\mathrm{d} \sigma_h}{\mathrm{d} y_h \mathrm{d}^2\boldsymbol{p}_T} = \frac{K}{(2\pi)^2} \frac{\sigma_0}{2} \int_{x_F}^1 \frac{\mathrm{d} z}{z^2}  \notag\\
& \times \sum_i x_1f_{i/p}(x_1,\mu^2) \widetilde{\mathcal{D}}_{F,A} \left(\frac{p_T}{z}, x_2\right) D_{h/i}(z,\mu^2)\,,
\end{align}
where $i$ runs over quark flavors and the gluon, and $\widetilde{\mathcal{D}}_{F,A}$ denotes the fundamental (adjoint) dipole for $i=q$ ($i=g$). The lower integration limit is given by $x_F = \frac{p_T}{\sqrt{s}} e^{y_h}$, which represents the minimum fraction of the projectile's momentum carried by the final-state hadron. The $K$-factor is introduced to account for missing higher-order corrections. The PDFs and FFs evolve according to DGLAP up to the factorization scale $\mu$. We will investigate different choices of this scale, as well as different fits to PDFs and FFs. The differential yield is obtained as $\mathrm{d} N_h/ (\mathrm{d} y_h \mathrm{d}^2\boldsymbol{p}_T) = \frac{1}{\sigma_{inel}} \mathrm{d} \sigma_h /(\mathrm{d} y_h \mathrm{d}^2\boldsymbol{p}_T) $, where $\sigma_{inel}$ represents the inelastic proton-proton cross-section.

\textbf{Numerical Setup.} For the initial condition of the BK equation, we parameterize the dipole amplitude using the $\text{MV}^\gamma$ model~\cite{McLerran:1997fk,Albacete:2010sy,Lappi:2013zma}:
\begin{equation}
    \mathcal{N}_F(r, x_0) = 1 - \exp \left[-\frac{(r^2 Q_{s0}^2)^\gamma}{4} \ln\left(\frac{1}{r \Lambda_{\text{QCD}}} +  e\right)\right]\,,
\end{equation}
where we have chosen $x_0 = 0.015$ to accommodate all data sets. $Q_{s0}^2, \gamma$ are parameters to be fitted, while $\Lambda_{\text{QCD}} = 0.241$~GeV. The adjoint dipole is obtained from the fundamental one in the large-$N_c$
limit as $\mathcal{D}_A = \mathcal{D}_F^2$, i.e.
$\mathcal{N}_A = 2\mathcal{N}_F - \mathcal{N}_F^2$~\cite{Lappi:2013zma}. For the evolution, we use Balitsky's prescription for the running coupling \cite{Balitsky:2006wa}. We also investigated the inclusion of kinematical constraints~\cite{Motyka:2009gi,Beuf:2014uia} to estimate NLO contributions. For the strong coupling, we employ the prescription used in~\cite{Beuf:2020dxl} for its smooth infrared behavior, which avoids unphysical dipole-tail fluctuations under Fourier transformation. We freeze the coupling in the infrared region at $\alpha_s \approx 0.76$. The fit parameter $C^2$ sets the running-coupling scale and evolution speed, while the fit parameter $\sigma_0/2$, the proton transverse area, sets the cross-section's overall normalization.

In the SIHP sector, we employ the \texttt{cteq6l1} set~\cite{Pumplin_2002} for PDFs and \texttt{NNFF1.0}~\cite{Bertone:2017tyb} for FFs. We choose the factorization scale as $\mu^2 = (p_T/z)^2 + Q_s^2$. This choice is motivated by NLO studies \cite{Chirilli:2011km,Chirilli:2012jd} that suggest that the natural choice is $\mu=c_0/r$ where $r$ is the dipole size. Since $p_T/z$ is Fourier conjugate to $r$, $c_0/r \sim p_T/z$. The saturation scale $Q_s^2$ acts as a natural infrared regulator, which we parametrize based on the GBW model~\cite{Golec-Biernat:1998zce}. Other choices of PDF and FF sets, and factorization scale are studied in the Appendix. In addition to the fit parameters $Q_{s0}^2, \gamma, \sigma_0/2$ and $C^2$, we include different $K$-factors at RHIC and the LHC, which account for missing higher-order corrections as discussed in the previous two sections. We consider only three light quark flavors ($u,d,s$) with $m=0.140$~GeV for both DIS and SIHP. In DIS, we perform the shift $x \rightarrow
x(1 + 4m^2/Q^2)$, where $m$ is the quark mass, which mainly affects the low-$Q^2$ region~\cite{Golec-Biernat:1998zce}.

Statistical uncertainties are determined using Hessian error estimation~\cite{Pumplin:2000vx,Pumplin:2001ct}, cross-validated against a Monte Carlo Bayesian analysis~\cite{Casuga:2023dcf, gelman_bayesian_2013}. We propagate FF uncertainties by averaging over the \texttt{NNFF1.0} replica ensemble and estimate theoretical errors through three-point scale variations, $\mu^2 \in [\mu^2/2, 2\mu^2]$. This approach ensures a robust determination of the correlation matrix.

\textbf{Results.} We performed a global analysis of HERA~\cite{H1:2015ubc} reduced cross-sections alongside forward SIHP at RHIC (BRAHMS~\cite{BRAHMS:2004xry} and STAR~\cite{STAR:2006dgg}) and LHCb ($\sqrt{s}=5$~\cite{LHCb:2021vww} and $13$~TeV~\cite{LHCb:2021abm}). For BRAHMS data, we adopt the inelastic cross-section $\sigma_{\text{inel}} = 41$~mb~\cite{BRAHMS:2004xry} to convert our cross-section into the measured yield. We apply kinematic cuts of $Q^2 \le 45$~GeV$^2$ for DIS and $p_T \geq 1$~GeV for SIHP. To isolate the impact of the SIHP data, we compared a six-parameter global fit with a four-parameter baseline restricted to DIS data. The analysis evaluates two evolution setups: the running-coupling BK (rcBK) equation with and without the kinematical constraint \cite{Ducloue:2019ezk} (see the Appendix for details).

\sbox{\tabboxone}{%
\begin{tabular}{l c c c c c c}
\hline\hline
Scheme & $Q_{s0}^2$ (GeV$^2$) & $\gamma$ & $C^2$ & $\sigma_0/2$ (mb) & $K$-RHIC & $K$-LHCb \\
\hline
\textbf{rcBK} &  0.1411(36) & 1.1521(77)& 10.3(1.1) & 17.26(28) & --- & --- \\
\hline
\textbf{kcBK} & 0.1084(27) &1.1135(73)& 1.276(68) & 19.69(31) & --- & --- \\
\hline
\textbf{rcBK} & 0.1428(33) & 1.1570(71) & 10.1(1.0) & 17.21(25) & 5.63(42) & 2.763(81)  \\
\hline
\textbf{kcBK} & 0.1107(27) & 1.1210(72) & 1.254(67) & 19.57(31) & 6.03(48) &3.049(94) \\
\hline\hline
\end{tabular}%
}
\edef\tabscaleone{\fpeval{\the\columnwidth/\the\wd\tabboxone}}

\begin{table}[H]
\centering
\caption{Best-fit parameters: rows 1–2 are the DIS-only baseline; the rest are the global fit.}
\label{tab:fit_results_params}
\resizebox{\columnwidth}{!}{\usebox{\tabboxone}}
\end{table}

\begin{table}[H]
\centering
\caption{$\chi^2$ values for various schemes. For the global fit,
$\text{d.o.f.} = N_{\text{points}} - N_{\text{params}}$, while for individual
datasets the d.o.f.\ equals the number of data points.}

\label{tab:fit_results_chi2}
\scalebox{\tabscaleone}{%
\begin{tabular}{l c c c c c}
\hline\hline
Scheme & $\chi^2_{\text{DIS}}/\textrm{d.o.f.}$ & $\chi^2_{\text{RHIC}}/\textrm{d.o.f.}$ & $\chi^2_{\text{LHC}}/\textrm{d.o.f.}$ & $\chi^2_{\text{tot}}/\textrm{d.o.f.}$ \\
\hline
\textbf{rcBK} & 0.97 & --- & --- & 0.975 \\
\hline
\textbf{kcBK} & 1.12 & --- & --- & 1.132 \\
\hline
\textbf{rcBK} & 0.97 & 0.32 & 0.48 & \textbf{0.850} \\
\hline
\textbf{kcBK} & 1.12 & 0.30 & 0.62 & 0.996 \\
\hline\hline
\end{tabular}%
}
\end{table}
Both evolution setups describe the DIS and SIHP data well (Tables~\ref{tab:fit_results_params}, \ref{tab:fit_results_chi2}), with rcBK giving the superior total fit ($\chisq/\text{d.o.f.}=0.850$). This preference comes from the DIS sector ($0.975$ versus $1.132$) and the LHCb spectra ($0.48$ versus $0.62$), while RHIC, with $\chisq/\text{d.o.f.}\approx0.3$ for both, does not discriminate between them given its larger uncertainties.
Interestingly, the kcBK-based fit prefers values of $\csq$ closer to
$e^{-2\gamma_E}$, which is the natural choice in the NLO calculation
\cite{Balitsky:2006wa}, whereas larger values of $\csq$ are required in the
rcBK-based fit to slow down the evolution. A similar pattern, in which
evolution equations that resum large transverse logarithms prefer coupling
scales close to the natural value, was observed in NLO fits to HERA
data~\cite{Beuf:2020dxl,Casuga:2025etc}.

\begin{figure}[hbt!]
    \centering
\includegraphics[width=0.75\linewidth]{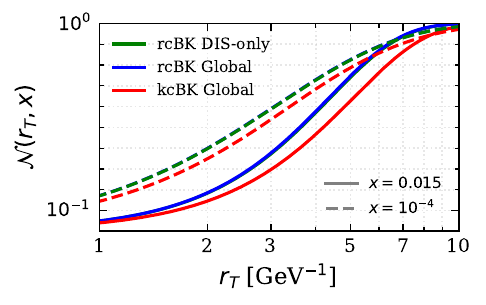}
    \caption{Dipole scattering amplitude $\mathcal{N}(r, x)$ at $x_0 = 0.015$ (solid) and $x = 10^{-4}$ (dashed).}
    \label{fig:initial}
\end{figure}

The interplay between the initial condition and the evolution also differs
between the two schemes. In the kcBK scheme, daughter dipoles smaller than
their parent are evaluated at shifted rapidities, which delays the growth of
the amplitude at short transverse distances and preserves the steepness of
its ultraviolet tail during the evolution. Smaller values of $\gam$ and
$\qso$ are then sufficient to describe the $x$ dependence of the data, while
the resulting reduction of the cross-section is compensated by a larger
$\sigma_0/2$. In the rcBK scheme, the evolution washes out the initial
steepness, and the fit compensates with larger $\gam$ and $\qso$ at $\xo$,
in addition to the larger $\csq$ noted above. Indeed, the two schemes yield
very similar evolved amplitudes in the kinematic region probed by the data,
as illustrated in Fig.~\ref{fig:initial} at $x=10^{-4}$.

\begin{figure}[H]
	\centering
    \includegraphics[width=\textwidth]{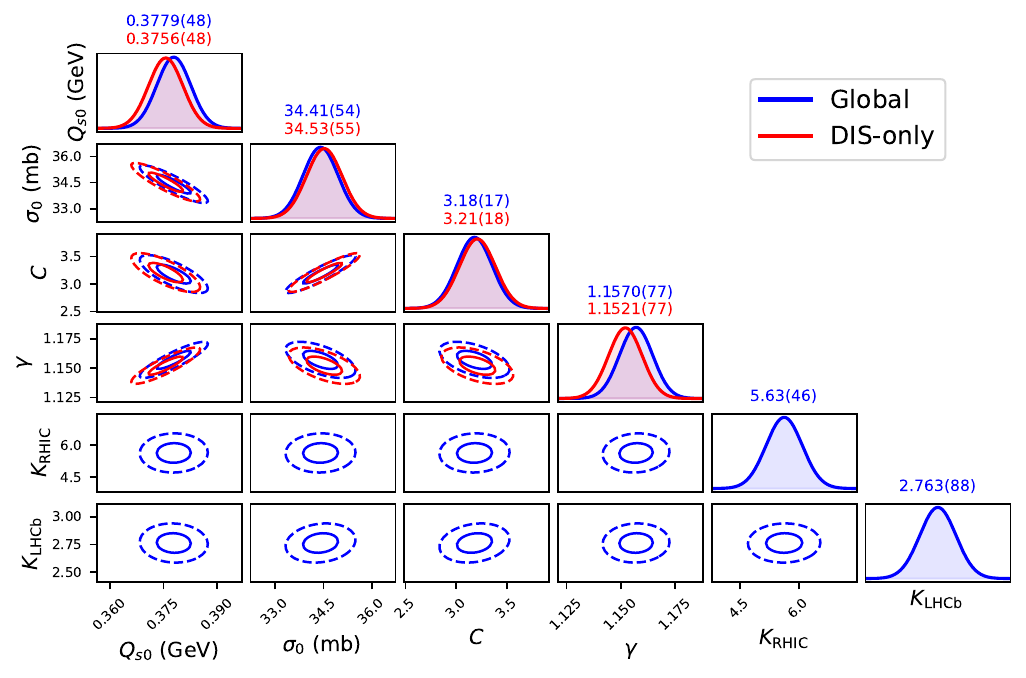}
    \begin{minipage}{\textwidth}
    \caption{Parameter correlations from the global and DIS-only fits using rcBK evolution. The off-diagonal panels display $1\sigma$ (solid) and
	$2\sigma$ (dashed) confidence ellipses.}
    \label{fig:correlation}
    \end{minipage}
\end{figure}

The extracted $K$ values differ between the schemes and depend
appreciably on the fragmentation-function set and the factorization scale
(see the Appendix). Importantly, the ordering $K_{\text{RHIC}} > K_{\text{LHCb}}$
is common to all setups we tested, with the exception of the anomalous
\texttt{NPC23} set (see the Appendix). The ratio $K_{\text{RHIC}}/K_{\text{LHCb}}$
depends on the fragmentation functions: it is approximately 2 for
\texttt{NNFF1.0} in both evolution schemes, 2.5 for \texttt{JAM19}, and 3
for \texttt{DSS} (see the Appendix). These values are consistent with NLO
threshold-resummed calculations in Ref.~\cite{Shi:2021hwx}, where the
resummed corrections are larger at RHIC than at the LHC. An analogous pattern is well known in collinear factorization, where LO
descriptions of hadron spectra require constant $K$-factors that decrease
systematically with increasing $\sqrt{s}$~\cite{Eskola:2002kv}. The residual
difference is due to higher-order corrections beyond threshold
logarithms and to the fragmentation-function and scale dependencies noted
above. Moreover, as shown in Fig.~\ref{fig:correlation}, the $K$-factors
are essentially uncorrelated with the remaining parameters, since the
normalization of the SIHP cross-section is absorbed by $K$ while $\sigma_0$
is determined by the large DIS dataset. The hadron spectra therefore
constrain the dipole amplitude only through their shape in $\pt$ and
rapidity. This makes the SIHP description a non-trivial test of the evolution: with a single rapidity-independent $K$ per collider, the
fit reproduces the LHCb spectra across six rapidity intervals and two
collision energies,  so the rapidity and energy dependence is captured by BK evolution itself, not by adjustable normalizations — unlike Ref.~\cite{Basso:2011fb}, which fits a separate normalization per rapidity bin.

\begin{figure}[hbt!]
    \centering
\includegraphics[width=0.85\linewidth]{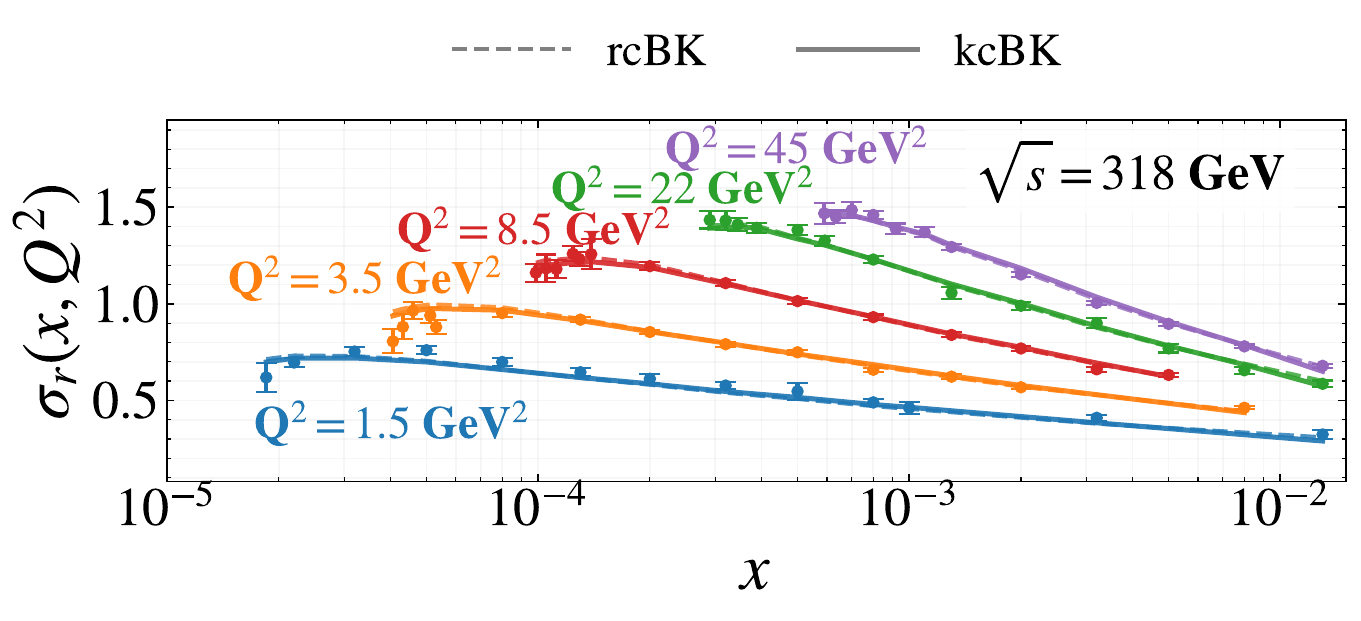}
    \caption{HERA reduced cross-section versus CGC calculations across $Q^2$ intervals in Bjorken $x$: dashed lines show rcBK evolution; solid lines show kcBK evolution.}
    \label{fig:dis}
\end{figure}

\begin{figure}[hbt!]
    \centering
\includegraphics[width=\linewidth]{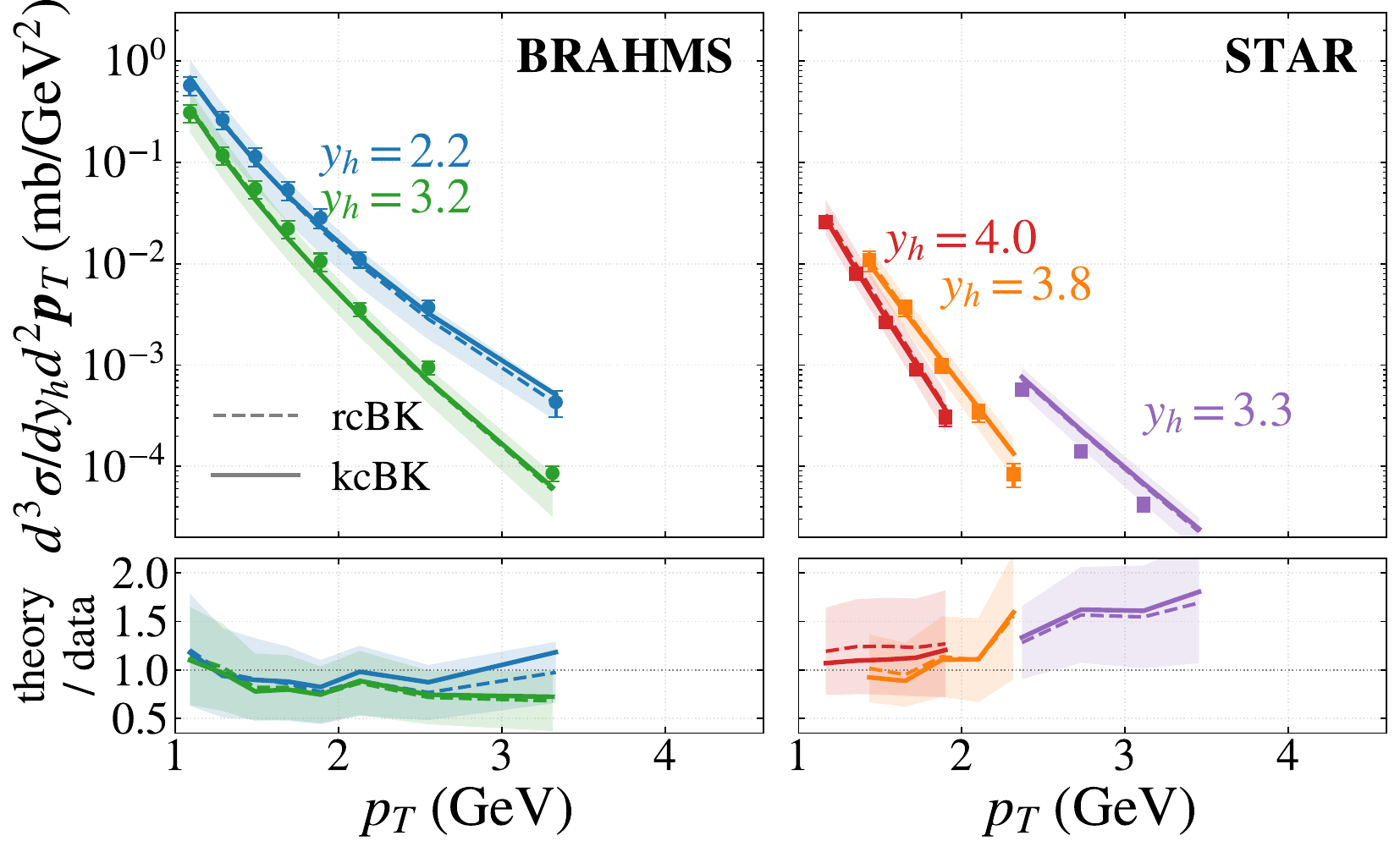}
    \caption{RHIC inclusive hadron production versus CGC calculations: dashed lines (bands) show rcBK evolution (uncertainty); solid lines show kcBK evolution.
}
    \label{fig:rhic}
\end{figure}

\begin{figure}[hbt!]
    \centering
\includegraphics[width=\linewidth]{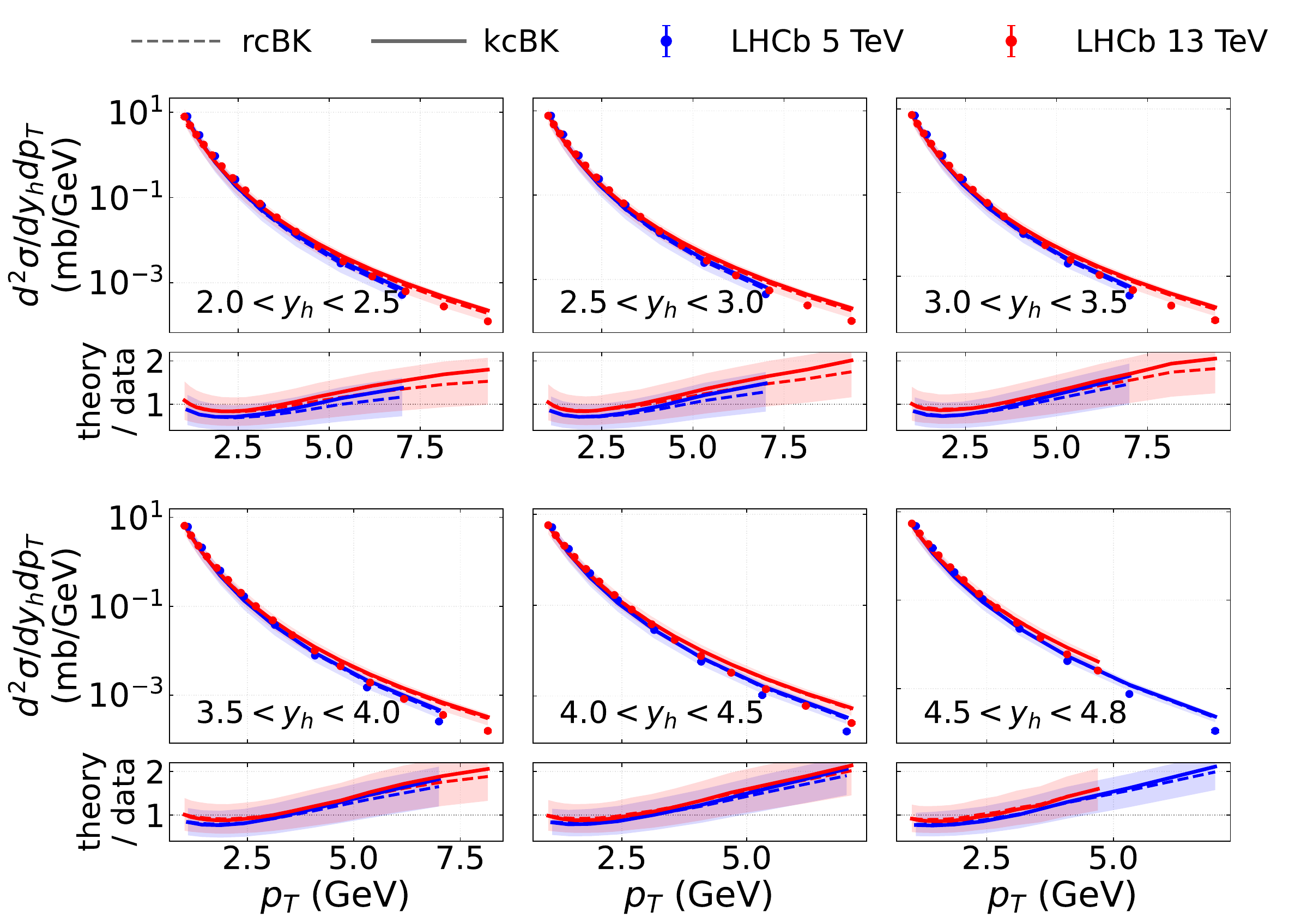}
    \caption{LHCb inclusive hadron production at $\sqrt{s}=5$ and 13 TeV versus CGC calculations: dashed lines (bands) show rcBK evolution (uncertainty); solid lines show kcBK evolution.}
    \label{fig:lhcb}
\end{figure}

Both schemes describe the reduced DIS cross-section well across four orders
of magnitude in $x$ (Fig.~\ref{fig:dis}). In the SIHP sector
(Figs.~\ref{fig:rhic} and \ref{fig:lhcb}), the two schemes (rcBK and kcBK) converge at the
highest rapidities, where the momentum fraction of the dense proton, $x_2$,
is smallest and the spectra are controlled by the evolved amplitude. They
mildly differ in the lowest-rapidity bins at $\pt \gtrsim 6$~GeV. In this region
the spectra probe the dipole at large transverse momenta $\pt/z$ and at
$x_2 \sim 10^{-4}$, only a few units of rapidity below $\xo$: the cross
section is thus sensitive to the tail of the amplitude at large momenta,
where the difference in the anomalous dimension $\gam$ between the two fits
has not yet been washed out by the evolution. The smaller $\gam$ preferred
by the kcBK fit corresponds to a slower fall-off of $\widetilde{\Ncal}$ at
large transverse momenta, and hence to a larger predicted cross-section.

The largest deviations occur at large $\pt$ and forward rapidities, where both schemes overshoot the data. Since a constant $K$-factor captures only the overall normalization, not its $\pt$ and rapidity dependence, this simplification likely drives part of the high-$\pt$ discrepancy: NLO threshold-resummed corrections shrink with $\pt$~\cite{Shi:2021hwx}, so a constant $K$-factor is expected to overshoot the data there.

The theoretical uncertainty in the SIHP sector is dominated
by the fragmentation functions, which we propagate through the
\texttt{NNFF1.0} replicas and which set the width of the bands in
Figs.~\ref{fig:rhic} and \ref{fig:lhcb}. In the low-$\pt$ region, the factorization scale based on the parent parton's
transverse momentum, $\mu \sim \pt/z$, supplemented with the saturation scale
as an infrared regulator, keeps the scale above the region where the LO gluon
fragmentation function develops unphysical negative values, and improves the
overall description of the data relative to the choice $\mu \sim \pt$ (see the Appendix).

We now quantify the impact of the SIHP data by comparing the DIS-only and
global fits. In our DIS-only rcBK fit, the parameters are close to those
of Ref.~\cite{Lappi:2013zma}. The smaller $\chisq$ reflects that the final HERA combination~\cite{H1:2015ubc} nearly doubles the
number of fitted points relative to the earlier
combination~\cite{H1:2009pze} used in Ref.~\cite{Lappi:2013zma}, and the
added points carry larger relative uncertainties. Relative to the DIS-only baseline, the global fit leaves the four
parameters unchanged within uncertainties and does not affect the DIS description ($\chisq_{\text{DIS}}/\text{d.o.f.} = 0.97$). This stability should be interpreted with care, since DIS contributes three times as many points as SIHP and FF uncertainties temper the SIHP $\chisq$ contribution. The DIS-only amplitude reproduces the $\pt$ and rapidity shapes of both colliders' hadron spectra, needing only two per-collider normalizations — consistent, given present FF-dominated uncertainties, with universality of the BK-evolved dipole amplitude.

The parameter correlations are studied in Fig.~\ref{fig:correlation}. A strong correlation exists between $\gam$ and $\qso$, since
they jointly determine the shape of the initial amplitude in the transition
region $(r \lesssim 1/Q_{s0})$, where an increase in $\qso$ must be
compensated by an increase in $\gam$. Similarly, $\csq$ and $\sigma_0$ are
correlated, as a slower evolution must be offset by a larger normalization, and since $\csq$ is anti-correlated with $\qso$, the pattern of the remaining correlations follows. We validated the Hessian uncertainties against an independent Monte Carlo Bayesian analysis (see Appendix). The two methods share no common approximation: the Hessian matrix is computed exactly at the minimum using automatic differentiation \cite{Cougoulic:2024jnd}, while the Bayesian analysis samples the posterior distribution directly from the likelihood. The agreement between the two, both for the individual parameter distributions and for the correlation ellipses, confirms the Gaussian behavior of the fitted parameters and validates the use of the much cheaper Hessian method for global analyses of this kind.

\onecolumngrid
\twocolumngrid

\textbf{Summary and Outlook.} In this letter we presented the first
simultaneous fit of the HERA reduced DIS cross-sections and forward single
inclusive hadron spectra at RHIC and the LHC in which the dipole amplitude is
obtained from BK evolution. We demonstrated that a good simultaneous
description of both observables, spanning $ep$ and $pp$ collisions from RHIC
to LHC energies, is achievable within the LO CGC framework supplemented by
per-collider $K$-factors. The rcBK setup provides the better overall
description, reaching a global $\chisq/\text{d.o.f.} = 0.85$, although the
kcBK setup favors values of $\csq$ closer to the natural choice
$e^{-2\gamma_E}$ expected from the NLO
calculation~\cite{Balitsky:2006wa}. The extracted $K$-factors are about a
factor of two larger at RHIC than at LHCb, consistent with threshold-resummed
one-loop studies of forward production~\cite{Shi:2021hwx}. The impact of the SIHP data on the extracted dipole amplitude is minimal:
the global-fit parameters coincide with the DIS-only baseline within
uncertainties. We find no tension between the two observables, which
supports the universality of the BK-evolved dipole amplitude. However, given the sizable dependence of the $K$-factors on
the fragmentation-function set and the factorization-scale prescription (see the Appendix), the constraining power of SIHP at this order should
not be overstated.

Natural extensions include proton-nucleus collisions, where the same framework predicts forward hadron spectra and nuclear modification factors $R_{pA}$ using the proton reference fixed by this fit. On the DIS side, adding the charm reduced cross-section would give a cleaner constraint, since the heavy-quark mass suppresses large-dipole contributions and thus the sensitivity to the non-perturbative region of the amplitude~\cite{Mantysaari:2018nng}. Beyond single-inclusive observables,
two-particle correlations such as photon-hadron, dihadron, and dijet
production offer direct access to the transverse momentum structure of the
target and stronger discrimination between saturation and higher-order
effects~\cite{Caucal:2025zkl}. Ultimately, we plan to extend this global analysis to NLO, for which the theoretical ingredients are now in place~\cite{Beuf:2020dxl,Hanninen:2022gje,Chirilli:2012jd,Liu:2019iml,Shi:2021hwx,Mantysaari:2023vfh,Boussarie:2025mzh,Boussarie:2025bpq,Casuga:2026xxt}, and stress-test the universality of the CGC effective theory.

\noindent{\bf Published data.} The grid with the dipole amplitude generated for the fitted parameters can be accessed at the RODBUK repository~\cite{rodbuk}.

\begin{acknowledgments}
\noindent{\bf Acknowledgments} We thank Florian Cougoulic, Heikki Mantysaari and Wenbin Zhao for valuable discussions related to this work. F.S. also acknowledges valuable discussions with Zhong-Bo Kang and Amanda Wei that motivated this project.  Numerical calculations were performed on the LUMI supercomputer under the time allocations: project\_465002091 (Calculating predictions for EIC physics) and project\_465002778 (Improving the precision of predictions for EIC physics). We gratefully acknowledge the Polish high-performance computing infrastructure PLGrid (HPC Center: ACK Cyfronet AGH) for providing computer facilities and support within the computational grant no. PLG/2024/017690. P.~K. and H.~L. are supported by the Polish National Science Center (NCN) grant No. 2022/46/E/ST2/00346. P.~K. thanks the EIC theory institute at BNL for its support and hospitality. F.S. is supported by the Laboratory Directed Research and Development of Brookhaven National Laboratory and RIKEN-BNL Research Center, as well as the National Science Foundation (NSF) within the framework of the JETSCAPE collaboration, under grant number OAC-2514008 (CSSI:C-SCAPE). F.S. also acknowledges the Saturated Glue (SURGE) Topical Theory Collaboration, funded by the U.S. Department of Energy, Office of Science, Office of Nuclear Physics. T.S.\ kindly acknowledges the support of the Polish National Science Center (NCN) Grant No.\,2021/43/D/ST2/03375.
\end{acknowledgments}

\bibliography{references}

\clearpage
\onecolumngrid
\begin{center}
    \textbf{\large Supplemental Material}
\end{center}
\vspace{1em}
\appendix
\setcounter{secnumdepth}{2}
\renewcommand{\thefigure}{A\arabic{figure}}
\renewcommand{\thetable}{A\arabic{table}}
\setcounter{figure}{0}
\setcounter{table}{0}

\section{Experimental data}
\label{sec:data}
Deep Inelastic Scattering (DIS) data from HERA---characterized by a small-$x$ kinematic range and a clean leptonic environment---provides an ideal baseline to probe the Color Glass Condensate (CGC) target structure (see Table~\ref{tab:data}). To ensure the applicability 
of the CGC framework where $Q^2 \gg Q_s^2$, we enforce an upper boundary of 
$Q^2 \le 45\text{ GeV}^2$ following Ref.~\cite{Lappi:2013zma}. On the other hand, where 
$Q^2 \rightarrow 0$, $Q_s^2$ becomes a semi-hard scale and prevents the non-perturbative 
effects of the strong coupling, meaning no lower cutoff is required. This kinematic selection 
yields a total of 532 data points. Within this regime, the published HERA neutral-current 
reduced cross-section ($\sigma_{r,\text{NC}}^+$) directly matches our theoretical fit quantity.

Conversely, the single inclusive hadron production (SIHP) sector can be measured with different final states across varying kinematic regimes. In the small-$x$ limit, forward rapidity is particularly suitable for probing the CGC. To preserve the validity of the hybrid formalism and ensure stable perturbative evolution, we enforce a 
lower boundary of $p_T \geq 1\text{ GeV}$. Within this kinematic regime, the SIHP sector includes a total of 172 data points, which is roughly one-third the size of the DIS dataset. While the BRAHMS and STAR collaborations operate at the same center-of-mass energy and similar rapidities, their differing final states 
($h^-$ versus $\pi^0$) lead to distinct behaviors in the global fit; this may manifest as different normalization $K$-factors. At higher energies, the LHCb measurements at $\sqrt{s} = 5$ and $13\text{ TeV}$ access an even smaller $x$-region and larger $p_T$, presenting a more stringent test for the framework. Despite these different collision energies, 
variations in the respective final-state observables compensate for the kinematic shifts, yielding experimental differential cross-sections of the same order of magnitude. Finally, to match the published experimental formats, our theoretical cross-section $\frac{d^3\sigma}{dy_h d^2\boldsymbol{p}_T}$ is scaled by the 
corresponding factors cataloged in Table~\ref{tab:factors}. For BRAHMS, the yield is 
obtained by dividing by the inelastic cross-section, for which we adopt 
$\sigma_{\text{inel}} = 41$~mb~\cite{BRAHMS:2004xry}. In addition to the factors listed, 
the units are matched to those of each measurement.

\begin{table}[H]
\caption{\label{tab:data}Summary of the experimental datasets included in the 
global analysis, including their specific observables, center-of-mass energies, 
number of data points, and kinematic coverage parameters.}
\begin{ruledtabular}
\begin{tabular}{lccccc}
Data Set & Process / Final State & $\sqrt{s}$ (GeV) & Points & $x_{\mathrm{Bj}}$ or $\eta_h$ & $p_T$ or $Q^2$ Range \\
\hline
DIS & $e^+ + p \to e^+ + X$ & 225--318 & 532 & $10^{-6}$--$10^{-2}$ & $0.045 \leq Q^2 \leq 45$~GeV$^2$ \\
BRAHMS & $p + p \to h^- + X$ & 200 & 16 & $2.2, \, 3.2$ & $1.09 \leq p_T \leq 3.33$~GeV \\
STAR & $p + p \to \pi^0 + X$ & 200 & 14 & $3.3, \, 3.8, \, 4.0$ & $1.167 \leq p_T \leq 3.45$~GeV \\
LHCb (5 TeV) & $p + p \to h^{\pm} + X$ & 5020 & 48 & $2.0 \leq \eta_h \leq 4.8$ & $1.105 \leq p_T \leq 7.0$~GeV \\
LHCb (13 TeV) & $p + p \to h^- + X$ & 13000 & 94 & $2.0 \leq \eta_h \leq 4.8$ & $1.1025 \leq p_T \leq 9.35$~GeV \\
\end{tabular}
\end{ruledtabular}
\end{table}

\begin{table}[H]
\caption{\label{tab:factors}Correspondence between the published experimental 
observables, units, and the specific conversion factors applied to our theoretical cross-section $\frac{d^3\sigma}{dy_h d^2\boldsymbol{p}_T}$.}
\begin{ruledtabular}
\begin{tabular}{lccccc}
 & DIS & BRAHMS & STAR & LHCb 5~TeV & LHCb 13~TeV \\
\hline
Experiment & $\sigma_{r,\text{NC}}^+$ & $\dfrac{d^3 N_h}{dy_h d^2 \boldsymbol{p}_T}$ & $E \dfrac{d^3\sigma}{d^3 \boldsymbol{p}} $&$\dfrac{d^2 \sigma}{dy_h dp_T}$ & $\dfrac{d^2 \sigma}{dy_h dp_T}$ \\[2.5ex]
Experimental Unit & Dimensionless & $\text{GeV}^{-2}$ & $\mu\text{b} \cdot \text{GeV}^{-2}$  &$\text{mb} \cdot \text{GeV}^{-1}$  & $\text{mb} \cdot \text{GeV}^{-1}$ \\[1ex]
Factor & 1 & $\dfrac{1}{\sigma_{\text{inel}}}$ & 1 & $2 \pi p_T$ & $2 \pi p_T$ \\
\end{tabular}
\end{ruledtabular}
\end{table}

\section{rcBK and kcBK evolution equations}
\label{sec:BK}

In this section, we provide the details of the BK equation used in this study. The LO BK equation reads \cite{Balitsky:1995ub,Balitsky:1998ya,Kovchegov:1999yj,Kovchegov:1999ua}
\begin{equation}
\frac{\partial S_{\bxt \byt}(\eta)}{\partial \eta} = \frac{\bar{\alpha}_s}{2 \pi} \int d^2 \bzt\   \mathcal{M}_{\bxt\,\byt\,\bzt} \big[ S_{\bxt\bzt}(\eta) S_{\bzt\byt}(\eta) - S_{\bxt\byt}(\eta)\big]\,,
\label{eq:BK}
\end{equation}
where $\bar{\alpha}_s= \frac{\alpha_s N_c}{\pi}$, and $\alpha_s$ is the strong coupling constant of QCD taken temporarily as a constant, $\eta=\ln(x_0/x)$ is the rapidity variable \footnote{Not to be confused with the hadron's pseudo-rapidity $\eta_h$.}, and $S_{\bxt \byt}(\eta)$ is the dipole amplitude describing the amplitude of scattering the pair of quark and antiquark of size $\brt = \bxt - \byt$ on a proton in the limit of very high energies.
The dipole kernel reads
\begin{equation}\label{eq:DipoleKernel}
    \mathcal{M}_{\bxt\,\byt\,\bzt} = \frac{(\bxt - \byt)^2}{(\bxt - \bzt)^2 (\bzt - \byt)^2}\,.
\end{equation}
The dipole amplitude depends only on the length $r=|\brt|$ and rapidity $\eta$, $S(r,\eta)$, because of the assumed translation and rotational invariance. The integral over $\bzt$ is estimated in radial coordinates, $r_z = |\bxt - \bzt|$ and $\phi = \angle (\bzt, \bxt)$. The  coordinate system's origin can be set at $\bxt$ without any loss of generality. Eq.~\eqref{eq:BK} rewritten in radial variables then reads 
\begin{equation}
\frac{\partial S(r,\eta)}{\partial \eta} = \frac{\bar{\alpha}_s}{2 \pi} \int d\phi \, dr_z\, r_z \, \frac{ r^2}{r_z^2 (r^2+r_z^2-2rr_z\cos\phi + \epsilon^2)}  \left[ S(r_z,\eta) \, S\left( \sqrt{r^2+r_z^2-2rr_z\cos\phi},\eta \right) - S(r,\eta)\right]\,.
\label{eq:BK_radial}
\end{equation}
This equation is UV safe, and we are allowed to regularize the pole in the kernel when $r^2+r_z^2-2rr_z\cos\phi \approx 0$ by a small $\epsilon$. The impact of $\epsilon$ on final results is negligible as the integral vanishes in this situation due to the vanishing color structure. We keep $\epsilon \sim 10^{-6}$ in our calculations.

\subsection{rcBK}

The rcBK equation is the LO BK equation that features the running of the coupling constant $\bar{\alpha}_s$. In order to avoid infrared problems, we implement the 
freezing of the coupling at large distances following  Ref.~\cite{Beuf:2020dxl},
\begin{equation}
    \bar{\alpha}_s(r) = \frac{N_c}{\pi} \frac{4 \pi}{\beta_0 \ln \Bigg[ \Big( \frac{\mu_0}{\Lambda_{\textrm{QCD}}} \Big)^{2/p} + \Big(\frac{2C}{r \Lambda_{\textrm{QCD}}} \Big)^{2/p}\Bigg]^p } \,,
\end{equation}
where we set
\begin{equation}
    \frac{\mu_0}{\Lambda_{\textrm{QCD}}} = 2.5 \,,
\end{equation}
which corresponds to a frozen value of about $\alpha_s(r \gg  \Lambda_{\textrm{QCD}}^{-1} ) \approx 0.76$. $p$ controls the smoothness of the transition between the running and frozen value. It is taken to be $0.2$  \cite{Beuf:2020dxl}. 
$C$ is a tunable, fitted parameter, while $\beta_0$ is the zeroth coefficient of the $\beta$-function in QCD, $\beta_0 = (11 N_c-2 N_f)/3$, where number of colors $N_c=3$ and number of quark flavors $N_f=3$. We favor this form of the running coupling as it leads to the smaller artifacts in the evolved dipole amplitude after Fourier transformation. Additionally, we follow Ref.~\cite{Balitsky:2006wa} in order to include higher-order corrections related to the running coupling. The full equation implemented in our software reads,
\begin{multline}
\frac{\partial S(r,\eta)}{\partial \eta} = \int d\phi \, dr_z\, r_z \,
\left[ \frac{\bar{\alpha}_s(r)}{2 \pi r^2_{z}} \left( \frac{r^2}{r^2_{zy} +\epsilon^2} + \frac{\bar{\alpha}_s(r_{z})}{\bar{\alpha}_s(r_{zy})}-1 + \frac{r^2_{z}}{r^2_{zy}+\epsilon^2} \left(\frac{\bar{\alpha}_s(r_{zy})}{\bar{\alpha}_s(r_{z})}-1 \right) \right) \right] 
\\
\times \big[ S(r_z, \eta) S(r_{zy}, \eta) - S(r, \eta)\big]\,,
\label{eq:BK_rc}
\end{multline}
where $r_{zy} = \sqrt{r^2+r_z^2-2r r_z\cos\phi}$\,.

\subsection{kcBK}

The kcBK equation derived in Ref.~\cite{Ducloue:2019ezk} which takes into account several additional physical effects, such as the running of the coupling constant with the energy scale, resummation of subleading corrections \cite{Andersson:1995ju,Kwiecinski:1996td, Motyka:2009gi}, etc. We quote the full form of the equation that is implemented in our software,
\begin{multline}
\frac{\partial S(r,\eta)}{\partial \eta} = \int d\phi \, dr_z\, r_z \,
\left[ \frac{\bar{\alpha}_s(r)}{2 \pi r^2_{z}} \left( \frac{r^2}{r^2_{zy} +\epsilon^2} + \frac{\bar{\alpha}_s(r_{z})}{\bar{\alpha}_s(r_{zy})}-1 + \frac{r^2_{z}}{r^2_{zy}+\epsilon^2} \left(\frac{\bar{\alpha}_s(r_{zy})}{\bar{\alpha}_s(r_{z})}-1 \right) \right) \right] 
\times \\
\times \big[ S(r_z, \eta - \delta_{r_z; r}) S(r_{zy}, \eta - \delta_{r_{zy}; r}) - S(r, \eta)\big]\,.
\label{eq:BK_kc}
\end{multline}
 The shifts in $\eta$ in the dipole amplitudes are given by $\delta_{r_z; r} = \max\Big\{0, 2 \log \frac{r}{r_{z}} \Big\}$ and similarly $\delta_{r_{zy}; r} = \max\Big\{0, 2 \log \frac{r}{r_{zy}} \Big\}$. 
Whenever the rapidity argument of the dipole amplitude is negative in the above square bracket, one uses the value of the dipole amplitude at the initial condition \cite{Ducloue:2019jmy}.

\section{Technical details}

In this section, we gather the technical details of our numerical framework. We describe the main features of the numerical setup for solving the BK equation in Sec.~\ref{sec:num_BK}. We employ and extend the approach described in \cite{Cougoulic:2024jnd} that allows to automatically estimate derivatives of the cross-sections with respect to the desired parameters. We provide a brief description of the method in Sec.~\ref{sec:num_AD}. One of the main benefits of this technique is the ability of exactly calculating the Hessian matrix at the minimum, and hence estimating the uncertainties. Since the cross-section for the single hadron inclusive production requires the dipole amplitude in momentum space, we propose in Sec.~\ref{sub:LFT} an algorithm that uses the standard, one-dimensional FFTW algorithm for calculating the Fourier transform of data on a logarithmic grid. The advantage of this algorithm is that, without overhead, it can be extended to dual numbers and hence easily offers access to automatic differentiation of the Fourier-transformed dipole amplitude.

\subsection{Solving the BK equation numerically}
\label{sec:num_BK}

The numerical solution of Eq.~\eqref{eq:BK_rc} or Eq.~\eqref{eq:BK_kc} can be obtained after discretizing the rapidity variable $\eta$. We tested two integration schemes: the Euler scheme and the simplest explicit Runge-Kutta scheme, and obtained comparable results. The variables $r$ and $r_z$ were discretized on a logarithmic grid \cite{Motyka:2009gi}. For all the integrations, we implemented the Simpson $\frac{3}{8}$ rule. Whenever the integration scheme required the value of the dipole amplitude between the nodes, that value was approximated using linear interpolation. The range for the $r$ and $r_z$ variables was set from $10^{-8}$ to 100 GeV$^{-1}$ and was discretized using 768 nodes. We checked that the discretization using 1024 nodes yielded equivalent results. In order to decrease the run time of the software, we parallelized the integrations using threads employing the \verb|openMP| framework, and we distributed the computations over several nodes using \verb|MPI| framework. Typically, the calculation of all the cross-sections (including the repetition over all Monte Carlo samples of the fragmentation functions for the SIHP cross-section) together with the full solution of the evolution equation required 4 minutes using 4 nodes built out of two AMD EPYC 7763 processors, with 64 cores each. 

\subsection{Minimization library}

We used the \verb|levmar| library \cite{lourakis04LM} to perform the optimization of the $\chi^2$. We interfaced the library to our parallel and distributed solver in such a way that process rank 0 was performing the minimization step and distributed the decision to the other processes.

The \verb|levmar| library offers two possible methods of estimating the gradients necessary to perform the iterative optimization step: either the derivatives are provided by the user, or they are estimated from finite differences by the library. Thanks to the automatic differentiation (described in the next subsection), we were able to provide the gradients of $\chi^2$ with respect to all parameters being optimized. This reduces the total numerical cost of the calculations. We also checked that the same solution is obtained if the library is used in the finite difference mode. 

\subsection{Automatic differentiation}
\label{sec:num_AD}

Automatic differentiation is a programming paradigm that offers access to semi-analytically evaluated derivatives over a user-specified set of parameters of the entire program without considerable modifications of the source code. We implemented our entire software using automatic differentiation, which allowed us to access the first (gradients) and second derivatives of $\chi^2$ with respect to the minimized parameters. Following Ref.~\cite{Cougoulic:2024jnd}, we introduced dual numbers and overloaded all operators needed for the solution of the BK equation. We extended \cite{Cougoulic:2024jnd} and managed to incorporate the Logarithmic Fourier Transform and the SIHP cross-section into the framework that allowed us to access first and second derivatives of the simultaneous $\chi^2$ for DIS and SIHP cross-sections.

\subsection{Logarithmic Fourier Transform} \label{sub:LFT}

The framework allows calculating the evolution of the dipole amplitude with the energy using the BK evolution equation. Often, the solution of the BK equation is obtained using a logarithmic grid, which improves the stability of the solution. From that, cross-sections for a large variety of processes can be estimated, some requiring the knowledge of the dipole amplitude in position space, and some in momentum space. The Logarithmic Fourier Transform (LFT) \cite{Talman}  provides an efficient way of finding the momentum representation of a function sampled in the position domain uniformly on a logarithmic scale. It has been widely adopted in the cosmological community \cite{Haines_Jones,Hamilton:1999uv,Hahn_2024,Rodriguez-Meza_2024}. In the context of transverse-momentum-dependent structure functions, the problem of Fourier transform was recently discussed in \cite{Kang:2019ctl,Ogata,Diehl:2024mmc}. 
The use of LFT not only improves the precision and reliability of the calculations but also is much faster than typical approaches. In the current application, our approach has the advantage that its implementation can be combined with a differentiable implementation of the BK solver using automatic differentiation and hence allows for propagating derivatives to the SIHP cross-section. Our implementation is based on the Python package \cite{python_lft} which was reimplemented in C++. More details on the analytic derivation of LFT and a performance analysis will be published elsewhere \cite{LFT}.

\begin{figure}[tb]
    \centering
    \includegraphics[width=\linewidth]{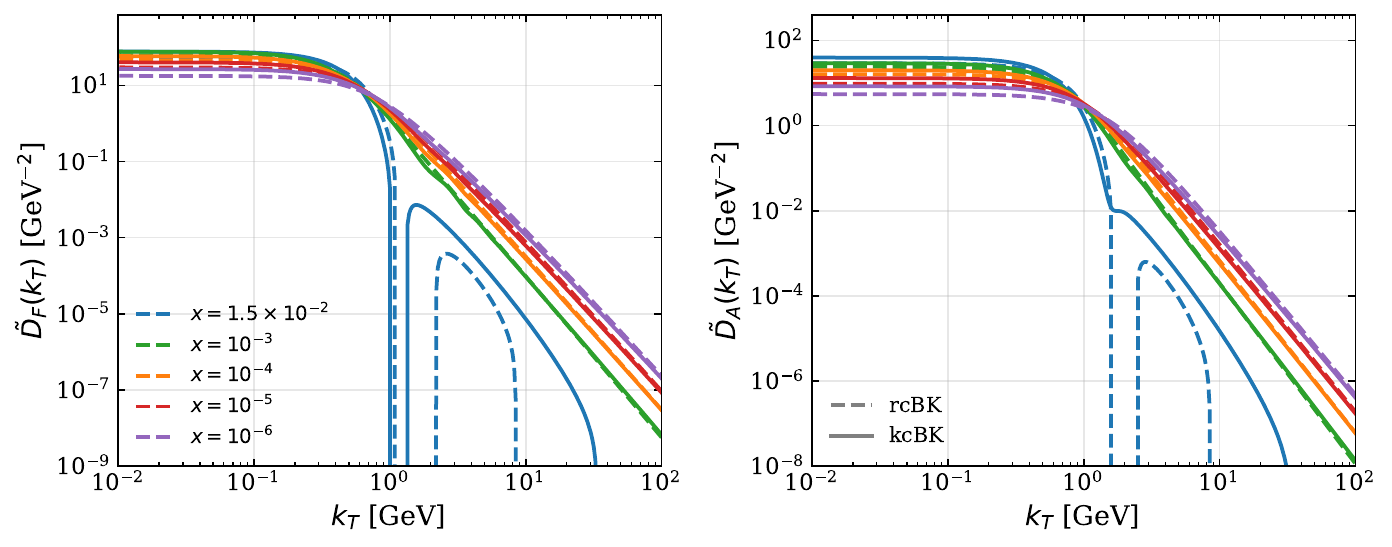}
    \caption{Fourier-transformed fundamental ($\tilde{D}_F$, left) and adjoint ($\tilde{D}_A$, right) dipole amplitudes for the global fit (\texttt{NNFF1.0}, scale $\mu^2=(p_T/z)^2+Q_s^2$), shown at several representative values of $x$ spanning the fitted range.}
    \label{fig:ft_dipole_appendix}
\end{figure}

\subsection{Estimation of the full Hessian matrix} \label{sub:hessian}

Usually, when the Hessian matrix is needed, it is approximated by an exterior product of gradient vectors. Automatic differentiation allows to estimate the exact Hessian matrix from the second derivatives with respect to the fitted parameters. This requires the implementation of the second derivatives of all functions required by the algorithm. In our case, we provide second derivatives of the \verb|log|, \verb|pow|, \verb|exp|, \verb|K_0|, and \verb|K_1| functions. After finding the optimal set of parameters, we reran our software with the calculation of second derivatives enabled, thus estimating the complete Hessian matrix at the minimum.

\section{Integration stability}\label{sec:integration_stability}
Under the Balitsky--Kovchegov (BK) equation at leading order (LO), the DIS cross-section is stable, in contrast to the SIHP sector, which introduces both technical and physical stability challenges. The most prominent issue stems from the Fourier transform of the dipole amplitude to momentum space; for large values of the anomalous dimension parameter $\gamma$, this transform is known to generate unphysical negative values at high $p_T$ and with values of $x_2$ near the initial value $x_0$ (Fig.~\ref{fig:ft_dipole_appendix}). In our fit parameter configuration, the extracted value of $\gamma = 1.16$ ensures that the negative regions of the integrand remain small and numerically negligible. To verify this, we performed a cross-check fit by explicitly setting the negative portions of the integrand to zero, which yielded no noticeable change in the final results.

Furthermore, the standard Fourier transform tends to produce spurious oscillations in the high-$p_T$ domain if the running coupling transitions abruptly into the infrared cutoff region. This artifact is regularized by adopting the smooth running coupling prescription from Ref.~\cite{Beuf:2020dxl}, and we follow the parameter values of $p$ and $\mu_0/\Lambda_{\textrm{QCD}}$ chosen in that study. Additionally, we implement the Logarithmic Fourier Transform framework, via
\begin{equation}
\tilde{D}_F(k) = -\frac{2\pi}{k} \int_0^\infty dr\, r\, J_1(k r)\, \mathcal{D}_F'(r)\,,
\qquad k>0\,,
\label{eq:LFT_derivative_appendixD}
\end{equation}
obtained by integrating the naive transform $\tilde{D}_F(k) = 2\pi\int_0^\infty dr\, r\, J_0(kr)\, \mathcal{D}_F(r)$ by parts. We find that transforming $\mathcal{D}_F'$ in place of $\mathcal{D}_F$ suppresses a spurious, non-decaying offset at large $k$ by roughly nine orders of magnitude on test functions with known closed-form transforms. On the model's initial dipole shape, $\mathcal{D}_F'$ continues to trace a smooth power-law decay nearly two decades beyond where $\mathcal{D}_F$ floors.

\begin{figure}[tb]

    \centering
    \begin{minipage}[b]{0.65\textwidth}
        \centering
        \includegraphics[width=\linewidth]{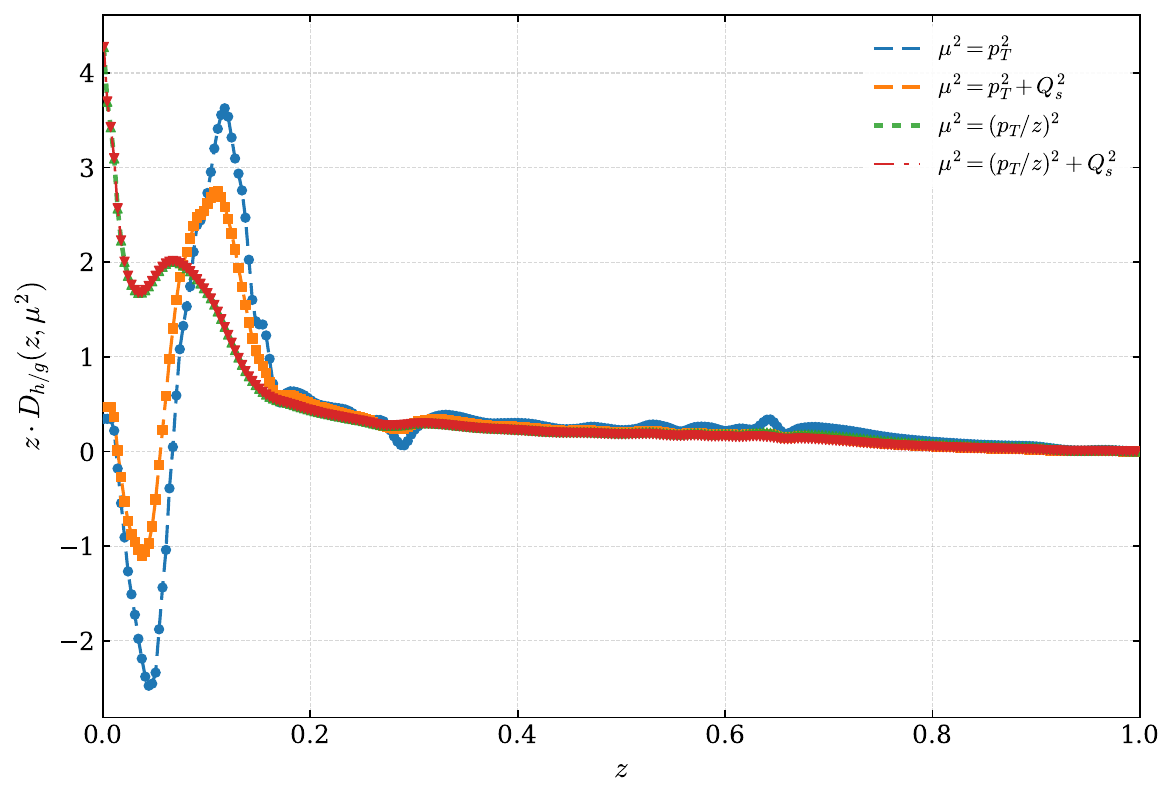}
        \vspace{2pt} 
        \centerline{\textbf{(a)}}
    \end{minipage}
    \hfill
    \begin{minipage}[b]{0.85\textwidth}
        \centering
        \includegraphics[width=\linewidth]{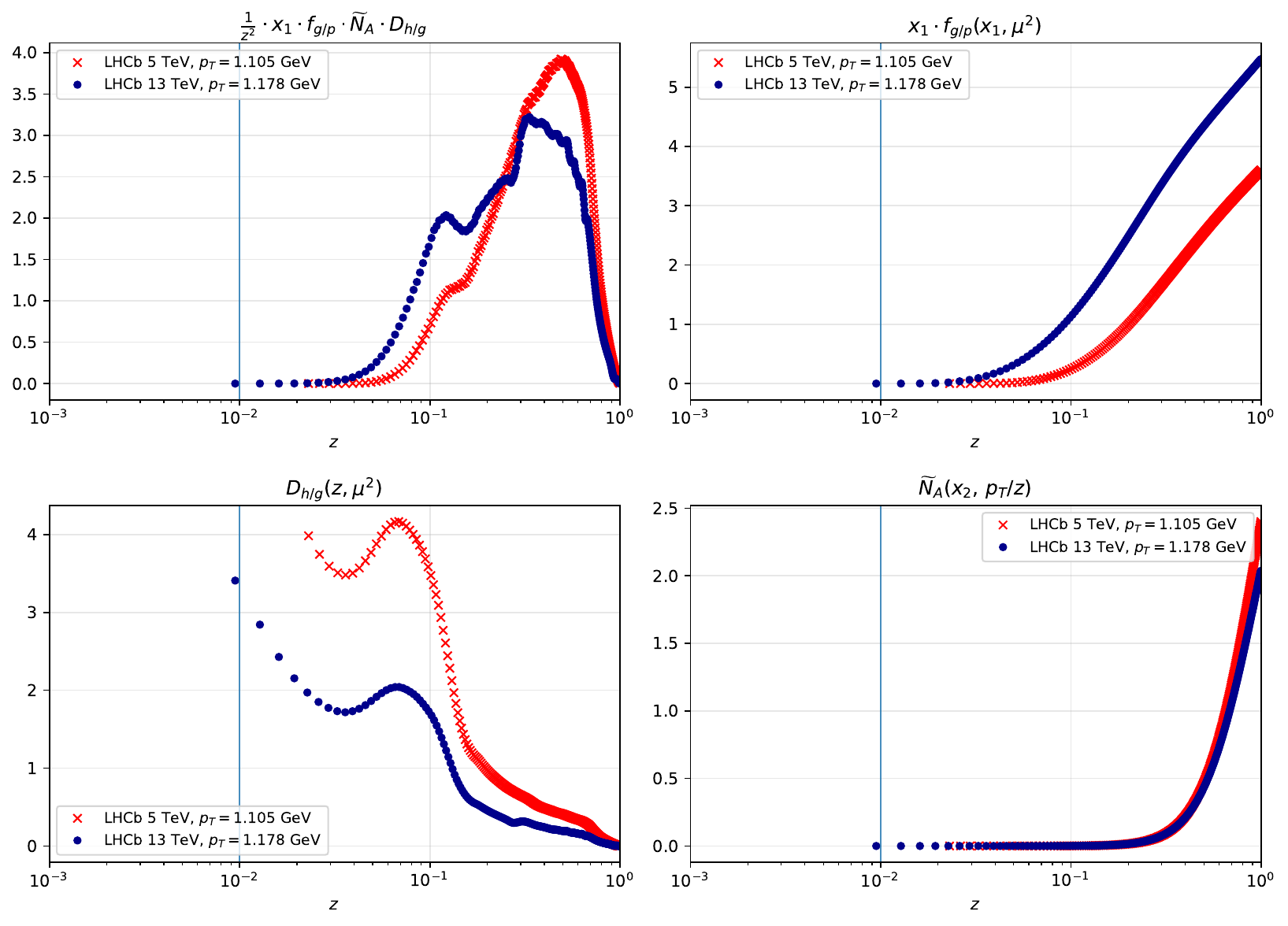}
        \vspace{2pt} 
        \centerline{\textbf{(b)}}
    \end{minipage}

    \caption{(a) Comparison of FF contributions under various factorization scale choices, demonstrating the regularization of negative values for LHCb 13 TeV kinematics at $\pt = 1.18$ GeV and $\eta_h = 2.25$. (b) Resulting stabilized SIHP integrand components plotted against $z$ at $\eta_h = 2.25$.}
    \label{fig:sihp_stability}
\end{figure}

Another source of negative integrand contributions involves the fragmentation functions (FFs) themselves, especially within next-to-leading order (NLO) frameworks. To avoid this behavior, our framework utilizes an LO FF baseline when it is available. While localized negative variations can still emerge at the boundaries of low $p_T$ and low rapidity where the system enters the deep saturation domain, implementing our specific factorization scale regularizes these regions. This improvement is demonstrated in Fig.~\ref{fig:sihp_stability}(a), which contrasts the FF behavior across different choices of the factorization scale, highlighting how our scale choice eliminates the negative contributions. Consequently, the resulting SIHP integrand achieves numerical stability, as illustrated by the 
component layouts in Fig.~\ref{fig:sihp_stability}(b).

\FloatBarrier

\section{Sensitivity to different fragmentation function (FF) and parton distribution function (PDF) sets} \label{sec:ff_choice}
\begin{figure}[H]
    \centering
    \begin{minipage}[b]{\linewidth}
        \centering
        \includegraphics[width=0.85\linewidth]{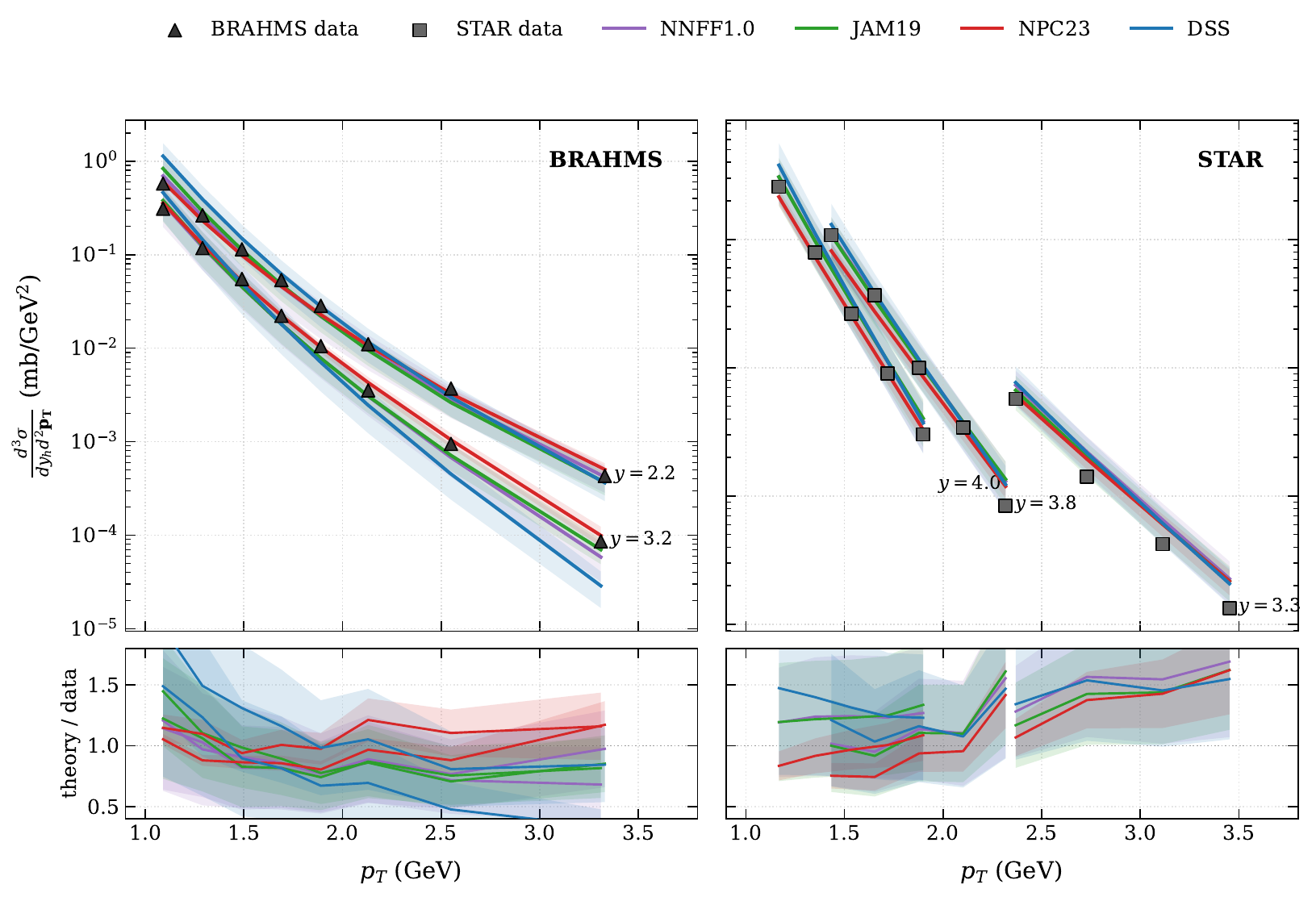}

        \vspace{2pt}
        \centerline{\textbf{(a)}}
    \end{minipage}
    \hfill
    \begin{minipage}[b]{\linewidth}
        \centering
        \includegraphics[width=0.98\linewidth]{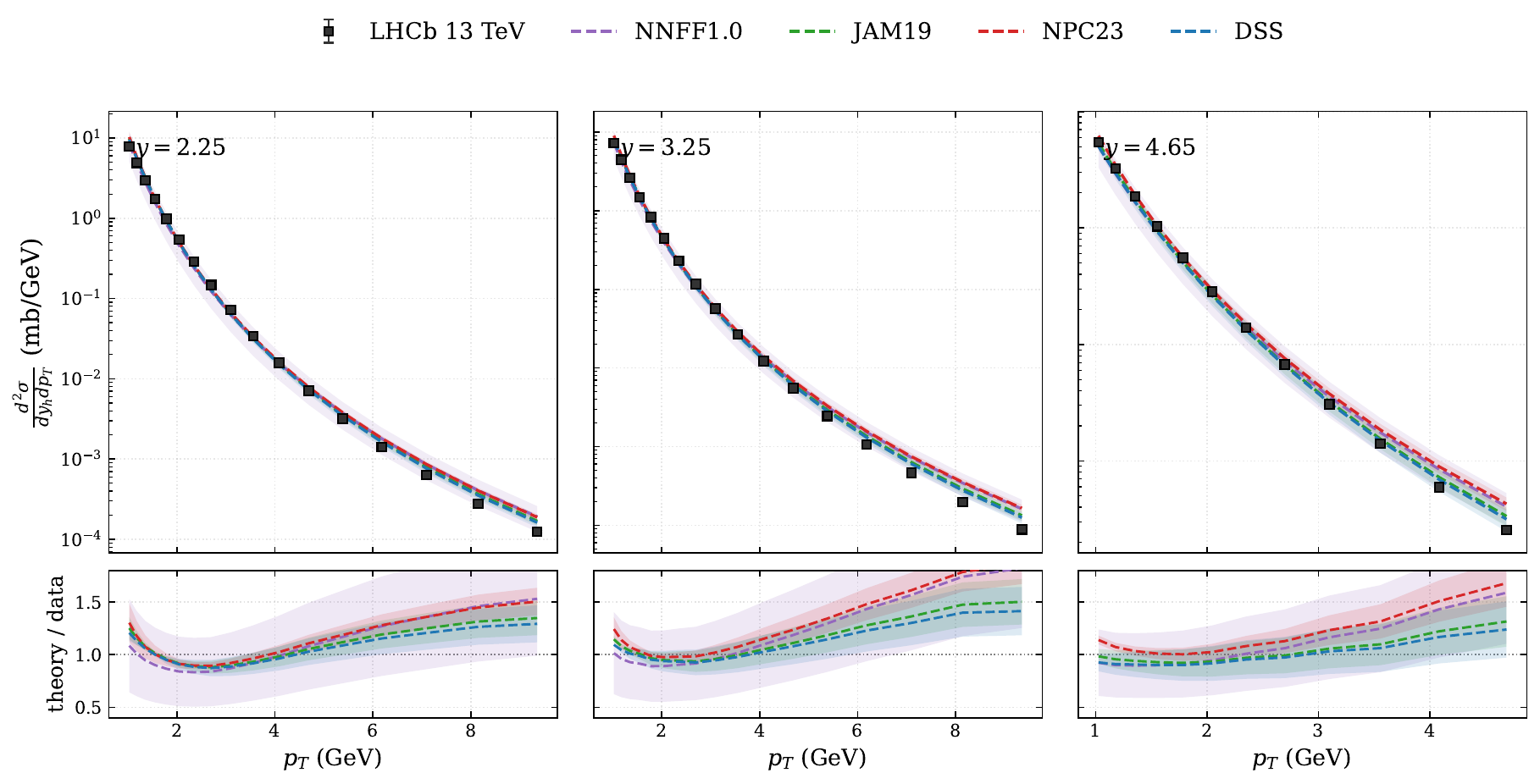}
        \vspace{2pt}
        \centerline{\textbf{(b)}}
    \end{minipage}
    \caption{Results for different sets of fragmentation functions (FFs) using the CTEQ PDF based on the best fit. Panel (a): RHIC data. Panel (b): selected experimental data points for LHCb 13 TeV data.}
    \label{fig:sigma_FF_Dependence}
\end{figure}

The results of the global fit for various FF sets are summarized in 
Table~\ref{tab:ffs} using the \texttt{cteq6l1} PDF. Corresponding plots with experimental data points are shown in Fig.~\ref{fig:sigma_FF_Dependence}. For the DSS set, we compute only the scale variation error due to the missing statistical uncertainty estimate.
The examined FF sets span distinct kinematic ranges in $Q$ and $x$ (Table~\ref{tab:ffs_sets}) and carry
significantly different error magnitudes. We compute the average percentage 
error relative to the SIHP differential cross-section across all observables, 
yielding values of 37\%, 16\%, 9.2\%, and 19\% for \texttt{NNFF1.0},
\texttt{JAM19}, \texttt{NPC23}, and \texttt{DSS}, respectively. Among the
baseline saturation parameters ($Q_{s0}^2$, $\sigma_0/2$, $C^2$, and
$\gamma$), the fit remains essentially unchanged across the different FF
sets, with \texttt{NPC23} the clear exception, shifting by up to $22\%$ in
$C^2$, $10\%$ in $Q_{s0}^2$, $4\%$ in $\sigma_0/2$, and $3\%$ in $\gamma$
relative to the \texttt{NNFF1.0} baseline. This behavior follows directly
from the tension between the SIHP spectra and the DIS-preferred description
at high $p_T$ discussed above. Since the FF theory error enters directly
into the SIHP $\chi^2$ weighting, \texttt{NPC23}, with the smallest error
among the four sets ($9.2\%$), gives the SIHP sector its largest statistical
weight in the fit, so that the high-$p_T$ tension is felt most strongly and
pulls the baseline parameters away from their DIS-only values. The other
three sets, with larger errors ($16$-$37\%$), dilute the same tension enough
that it leaves little imprint on the fit.

To test this, we consider the \texttt{NPC23} set with a transverse-momentum
cutoff ($p_T \leq 6$~GeV, sixth row), which removes part of the high-$p_T$
SIHP data responsible for the tension. Under this constraint, all four
baseline parameters shift back close to the \texttt{NNFF1.0} reference
values in the first row. Removing the tension by hand thus reproduces the
same effect as diluting it through a larger theory error, confirming that it
is specifically the
high-$p_T$ tension -- and how much statistical weight it is given -- rather
than the choice of input FF set itself, that drives the shift in the
baseline parameters observed for \texttt{NPC23}.

Conversely, the normalization $K$-factors exhibit a clear dependency on the 
chosen FF set. Even with the $p_T$ cut implemented in \texttt{NPC23}, the 
resulting $K$-factors vary by up to a factor of two or three compared to the 
\texttt{NNFF1.0} reference baseline. Furthermore, the number of independent 
normalization parameters required to optimize the fit depends on the chosen 
framework. The \texttt{NNFF1.0} set is well-described using a single
independent $K$-factor for RHIC, as illustrated by its collapsed $K$-factor
columns; the same holds for \texttt{JAM19} and \texttt{DSS}. \texttt{NPC23}
is the only set that requires two separate $K$-factors, $K_{\text{BRAHMS}}$
and $K_{\text{STAR}}$, despite BRAHMS and STAR operating at identical
collision energies. On the other hand, a single $K$-factor remains
sufficient for the LHCb data. This may stem from the fact that the LHCb 
datasets share similar final states ($h^-$ and $h^\pm$) that utilize two FF 
sets related by a factor of two. This contrast stands unlike the case of 
BRAHMS and STAR, where completely different hadron species ($\pi^0$ and $h^-$) 
are measured, necessitating the use of distinct, uncorrelated FF sets. This 
trend suggests that while the baseline geometric parameters remain stable in 
the absence of kinematic tension, the overall normalization scale is strongly 
correlated with the chosen fragmentation functions.

Additionally, we perform a cross-check calculation for the \texttt{DSS} set 
by adopting the model parameters from~\cite{Lappi:2013zma}. The extracted 
$K$-factors are close to $2.5$, which indicates a reasonable agreement with the 
values reported in that study. 

Finally, we checked that comparing \texttt{cteq6l1}~\cite{Pumplin_2002} and \texttt{mstw}~\cite{mstwpdf} PDF sets, the overall 
results do not exhibit a significant sensitivity to the specific choice of the input PDF set.

\begin{table}[htbp]
\centering
\begin{tabular}{|c|c|c|}
    \hline
    FFs & $Q \text{ range}$ (GeV) & $x \text{ limits}$  \\ \hline
    
    NNFF1.0 (LO) & $1-10000$ & $0.01 - 1$  \\ \hline
    
    JAM19 (NLO) & $1.14 - 316$ & $10^{-6} - 1$  \\ \hline
    
    DSS (LO) & $1 - 316.23$ & $0.01 - 1$  \\ 
    
    \hline 
    \hline

    NPC23 (NLO) & $4 -4000$ & $0.003 - 1$  \\ \hline

\end{tabular}
\caption{Kinematical boundaries of the fragmentation function sets tested in this work. We treat the NPC23 set separately because of a higher $Q$ range. As explained in the text and in Table~\ref{tab:ffs}, this FF set leads to larger $K$-factors than the other sets.}
\label{tab:ffs_sets}
\end{table}

\begin{table}[htbp]
\centering
\begin{tabular}{|c|c|c|c|c|c|c|c|c|c|c|}
    \hline
    FFs & $Q_{s0}^2$ ($\text{GeV}^2$) & $\sigma_0/2$ ($\text{mb}$) & $C^2$ & $\gamma$ & 
    $K_{\text{BRAHMS}}$ & $K_{\text{STAR}}$ & $K_{\text{LHCb5}}$ & $K_{\text{LHCb13}}$ &$\color{blue}{\chi^2_{\text{DIS}}/\text{d.o.f.}}$ & $\chi^2/\text{d.o.f.}$ \\ \hline
    
    \multirow{2}{*}{\texttt{NNFF1.0} (LO)} & \multirow{2}{*}{0.143} & \multirow{2}{*}{17.206} & \multirow{2}{*}{10.117} & \multirow{2}{*}{1.157} &
    \multicolumn{2}{c|}{5.630} & \multicolumn{2}{c|}{2.763} & \multirow{2}{*}{\color{blue}{0.969}} & \multirow{2}{*}{0.850} \\ \cline{6-9}
    &&&&& \color{blue}{0.245} & \color{blue}{0.396} & \color{blue}{0.671} & \color{blue}{0.383} && \\ \hline

    \multirow{2}{*}{\texttt{JAM19} (NLO)} & \multirow{2}{*}{0.149} & \multirow{2}{*}{16.688} & \multirow{2}{*}{8.342} & \multirow{2}{*}{1.164} &
    \multicolumn{2}{c|}{4.852} & \multicolumn{2}{c|}{1.948} & \multirow{2}{*}{\color{blue}{0.977}} & \multirow{2}{*}{1.138} \\ \cline{6-9}
    &&&&& \color{blue}{0.336} & \color{blue}{0.388} & \color{blue}{2.171} & \color{blue}{1.703} && \\ \hline
    
    \multirow{2}{*}{\texttt{DSS} (LO)} &  \multirow{2}{*}{0.147} & \multirow{2}{*}{16.820} & \multirow{2}{*}{8.714} & \multirow{2}{*}{1.161} &
    \multicolumn{2}{c|}{4.803} & \multicolumn{2}{c|}{1.638} & \multirow{2}{*}{\color{blue}{0.973}} & \multirow{2}{*}{1.005} \\ \cline{6-9}
    &&&&& \color{blue}{1.115} & \color{blue}{0.405} & \color{blue}{1.255} & \color{blue}{1.063} && \\
    
    \hline
    \hline

    \multirow{2}{*}{\texttt{NPC23} (NLO)} & \multirow{2}{*}{0.157} & \multirow{2}{*}{16.474} & \multirow{2}{*}{7.913} & \multirow{2}{*}{1.190} &
    15.531 & 5.059 & \multicolumn{2}{c|}{5.744} & \multirow{2}{*}{\color{blue}{1.010}} & \multirow{2}{*}{1.922} \\ \cline{6-9}
    &&&&& \color{blue}{0.281} & \color{blue}{0.785} & \color{blue}{7.538} & \color{blue}{4.520} && \\ \hline
    
    \multirow{2}{*}{\begin{tabular}{c}\texttt{NPC23} (NLO)\\$p_T \leq 6$ GeV\end{tabular}} &  \multirow{2}{*}{0.143} & \multirow{2}{*}{17.127} & \multirow{2}{*}{10.133} & \multirow{2}{*}{1.155} &
    14.746 & 5.090 & \multicolumn{2}{c|}{5.840} & \multirow{2}{*}{\color{blue}{0.970}} & \multirow{2}{*}{1.507} \\ \cline{6-9}
    &&&&& \color{blue}{0.265} & \color{blue}{0.952} & \color{blue}{6.061} & \color{blue}{2.956} && \\ \hline

\end{tabular}
\caption{Results for different choices of the fragmentation function sets. The results obtained with the NPC23 set differ significantly from the other choices; in particular, the values of the $K$-factors are larger.}
\label{tab:ffs}
\end{table}

\section{Choice of the factorization scale}
\label{sec:scale_choice}
The evaluation of PDFs and FFs within the SIHP sector requires a proper factorization scale $\mu$. While conventional physical choices include the transverse momentum of the parent parton ($p_T/z$) 
or the produced hadron ($p_T$),  the latter choice is sub-optimal relative to $p_T/z$ in the low-$p_T$ kinematic domain. In this soft region, 
a pure $p_T$ scale exhibits limited sensitivity to the onset of gluon 
saturation dynamics and weakens the stability of perturbative theory. 
To regularize this behavior, a phenomenological prescription is 
$\mu^2 = \alpha^2(\mu_{\min}^2 + p_T^2)$~\cite{Shi:2021hwx}, which introduces 
parameters $\alpha$ and $\mu_{\min}$ that must be tuned manually.

In this work, we implement a physically motivated scale definition:
\begin{equation}
\mu^2 = \left(\frac{p_T}{z}\right)^2 + Q_s^2(x),
\label{eq:default_scale}
\end{equation}
where $Q_s(x)$ represents the dynamical saturation scale. This formulation effectively 
demarcates the transition between the collinear perturbative regime and the saturation 
domain. We adopt the standard parameterization from the Golec-Biernat--Wüsthoff (GBW) 
saturation model~\cite{Golec-Biernat:1998zce}, given by $Q_s^2(x) = Q_0^2 (x_0^{\text{GBW}}/x)^\lambda$,
with parameters fixed to $Q_0^2 = 1~\text{GeV}^2$, $x_0^{\text{GBW}} = 3.04 \times 10^{-4}$,
and $\lambda = 0.288$~\cite{Golec-Biernat:1998zce}.

The comparison of different scale choices is given in Table~\ref{tab:global_fit_rcbk}. Our default scale choice, Eq.~\eqref{eq:default_scale}, leads to the smallest $\chi^2$ value.
Furthermore, our results reveal that while the baseline model parameters ($Q_{s0}^2$, $\sigma_0/2$, $C^2$, and $\gamma$) remain less sensitive to the choice of factorization scale, the extracted $K$-factors exhibit a strong dependence on the chosen 
scale prescription.
\begin{table}[H]
\centering
\begin{tabular}{|c|c|c|c|c|c|c|c|c|c|c|}
\hline
$\mu^2$ & $Q_{s0}^2$ (GeV$^2$) & $\sigma_0/2$ (mb) & $C^2$ & $\gamma$ & \multicolumn{2}{c|}{$K_{\text{RHIC}}$ / $\color{blue}{\chi^2_{\text{RHIC}}}/\text{d.o.f.}$} & \multicolumn{2}{c|}{$K_{\text{LHCb}}$ / $\color{blue}{\chi^2_{\text{LHCb}}}/\text{d.o.f.}$} &$\color{blue}{\chi^2_{\text{DIS}}}/\text{d.o.f.}$ & $\chi^2/\text{d.o.f.}$ \\ \cline{6-9}
& & & & & \scriptsize BRAHMS & \scriptsize STAR & \scriptsize 5 TeV & \scriptsize 13 TeV & & \\ \hline

\parbox[c]{2.2cm}{\vspace{0.15cm}\centering $\left(\frac{p_T}{z}\right)^2 + Q_s^2$\vspace{0.15cm}} & 0.143 & 17.206 & 10.117 & 1.157 & \multicolumn{2}{c|}{\scriptsize $K$ = 5.630} & \multicolumn{2}{c|}{\scriptsize $K$ = 2.763} & {\color{blue}0.969} & \textbf{0.850} \\ \cline{6-9}
&&&&& {\color{blue}\scriptsize 0.245} & {\color{blue}\scriptsize 0.396} & {\color{blue}\scriptsize 0.671} & {\color{blue}\scriptsize 0.383} && \\
\hline

\parbox[c]{2.2cm}{\vspace{0.15cm}\centering $4(4+p_T^2)$\vspace{0.15cm}} & 0.144 & 17.152 & 9.922 & 1.160 & \multicolumn{2}{c|}{\scriptsize $K$ = 9.048} & \multicolumn{2}{c|}{\scriptsize $K$ = 2.901} & {\color{blue}0.970} & 0.883 \\ \cline{6-9}
&&&&& {\color{blue}\scriptsize 0.244} & {\color{blue}\scriptsize 0.489} & {\color{blue}\scriptsize 0.920} & {\color{blue}\scriptsize 0.484} && \\
\hline

\parbox[c]{2.2cm}{\vspace{0.15cm}\centering $9(4+p_T^2)$\vspace{0.15cm}} & 0.144 & 17.127 & 9.831 & 1.161 & \multicolumn{2}{c|}{\scriptsize $K$ = 11.595} & \multicolumn{2}{c|}{\scriptsize $K$ = 3.209} & {\color{blue}0.970} & 0.899 \\ \cline{6-9}
&&&&& {\color{blue}\scriptsize 0.280} & {\color{blue}\scriptsize 0.532} & {\color{blue}\scriptsize 1.077} & {\color{blue}\scriptsize 0.507} && \\
\hline

\parbox[c]{2.2cm}{\vspace{0.15cm}\centering $\pt ^2 + Q_s^2$\vspace{0.15cm}} & 0.143 & 17.195 & 10.090 & 1.158 & \multicolumn{2}{c|}{\scriptsize $K$ = 4.403} & \multicolumn{2}{c|}{\scriptsize $K$ = 2.605} & {\color{blue}0.969} & 0.897 \\ \cline{6-9}

&&&&& {\color{blue}\scriptsize 0.199} & {\color{blue}\scriptsize 0.293} & {\color{blue}\scriptsize 0.870} & {\color{blue}\scriptsize 0.653} && \\
\hline
\end{tabular}
\caption{Global fit minimization results utilizing the rcBK evolution equation. PDF and FF sets are \texttt{cteq6l1} and \texttt{NNFF1.0}, respectively.}

\label{tab:global_fit_rcbk}
\end{table}

\section{Uncertainties determination}

In this section, we describe two independent methods of obtaining uncertainties of the fitted parameters. In Sec.~\ref{sec:hessian}, we provide the details of extracting the covariance matrix from the Hessian matrix. The uncertainties obtained in this way are assumed to be normally distributed as the Hessian only captures the quadratic shape of this minimized $\chi^2$ function. On the contrary, in Sec.~\ref{sec:MC}, we describe the Monte Carlo Bayesian inference approach, which is valid generally; however comes with a much higher cost than the Hessian method. In Sec.~\ref{sec:MC_vs_hessian}, we compare the two methods, cross-checking both methods and thus confirming the Gaussian distribution of the fitted parameters.

\subsection{Hessian matrix}
\label{sec:hessian}
We obtain the full Hessian matrix $H$ at the minimum thanks to automatic differentiation. Hence, we are not making the approximation where the Hessian matrix is calculated from the external product of gradient vectors (commonly known as the Gauss-Newton approximation), allowing us to preserve non-linear structural information near the minimum. The covariance matrix $\Sigma$ is related to the Hessian matrix through,
\begin{equation}
    \Sigma = (H/2)^{-1}, \quad H_{ij} = \dfrac{\partial^2 \chi^2}{\partial \theta_i \partial\theta_j}
\end{equation}
For a general multi-parameter system, the joint uncertainty of any two parameters is isolated by extracting the corresponding $2\times2$ sub-block of the full covariance matrix. For clarity of presentation, we assume that the minimization was performed for two specific parameters denoted $x$ and $y$. In that case, the resulting $2\times2$ covariance matrix $\Sigma$ is defined as:
\begin{equation}
\Sigma = \begin{pmatrix} 
\sigma_x^2 & \rho \sigma_x \sigma_y \\ 
\rho \sigma_x \sigma_y & \sigma_y^2 
\end{pmatrix}\,,
\end{equation}
where $\sigma_x$ and $\sigma_y$ denote the standard errors of the parameters $x$ and $y$, 
respectively, and $\rho = \frac{\text{cov}(x,y)}{\sigma_x \sigma_y}$ is the Pearson correlation 
coefficient, which acts as a dimensionless measure of linear dependence bounded by 
$\rho \in [-1, 1]$. This matrix can be factored into a diagonal scaling matrix $Q$ and a 
normalized correlation matrix to completely decouple the individual parameter scales 
from the pure correlation structure:
\begin{equation}
\begin{pmatrix} 
\sigma_x^2 & \rho \sigma_x \sigma_y \\ 
\rho \sigma_x \sigma_y & \sigma_y^2 
\end{pmatrix} = Q \begin{pmatrix} 
1 & \rho \\ 
\rho  & 1 
\end{pmatrix} Q\, \qquad \textrm{where} \qquad Q = \begin{pmatrix}  
\sigma_x & 0 \\ 
0 & \sigma_y 
\end{pmatrix}\,.
\end{equation}

An eigenvalue decomposition of the correlation matrix isolates the invariant geometric 
properties of the correlation field, yielding:
\begin{equation}
\begin{pmatrix} 
1 & \rho \\ 
\rho  & 1 
\end{pmatrix} = C \begin{pmatrix} 
1 + \rho & 0 \\ 
0  & 1 - \rho 
\end{pmatrix} C^T\,, \qquad \textrm{where} \qquad C = \begin{pmatrix} 
\frac{1}{\sqrt{2}} & -\frac{1}{\sqrt{2}} \\ 
\frac{1}{\sqrt{2}}  & \frac{1}{\sqrt{2}} 
\end{pmatrix}\,,
\end{equation}
represents a $45^\circ$ rotation matrix. Because the diagonal elements of the core 
correlation matrix are identical, these eigenvectors are stationary and universal, 
meaning the principal axes of any two-parameter correlation system are always 
oriented exactly at $45^\circ$ regardless of the strength of $\rho$. The eigenvalues, 
$1+\rho$ and $1-\rho$, explicitly show how the total variance is redistributed along 
these major and minor axes.

To construct the confidence ellipse contours, an initial ellipse with semi-axes lengths 
$(\sqrt{1+\rho}, \sqrt{1-\rho})$ is generated in the normalized eigenspace. The rotation 
matrix $C$ is then applied to orient the ellipse by $45^\circ$ along the true empirical 
correlation axes, followed by a final linear transformation via the scaling matrix 
$Q$ to map it from the canonical space back into the physical parameter space.

\subsection{Monte Carlo Bayesian inference}
\label{sec:MC}

In the context of performing fits, Monte Carlo Bayesian inference allows for generating the posterior distribution of fitted parameters $\theta$. In the discussed application, $\theta = \{ Q_{s0}^2, \sigma_0/2, C^2, \gamma, \dots\}$. Using the Bayes theorem and assuming a uniform prior distribution for the fit parameters, the posterior is directly proportional to the likelihood, given by
\begin{equation}
    L({\theta}) = \exp(-\chi^2(\theta)/2) \,.
\end{equation}
Typically, sampling from $L(\theta)$ is not possible, as in the present context, evaluating $\chi^2(\theta)$ requires solving the BK equation and convoluting the resulting dipole amplitude into appropriate cross-sections. Hence, in order to sample parameters $\theta$ from the posterior, one constructs a Markov chain with an equilibrium probability distribution given by the likelihood. In the simplest setup, a Metropolis-type of algorithm can be set up, which proposes small changes to the parameters $\theta_{n+1} = \theta_n + \delta$. The proposed parameters are accepted with the probability
\begin{equation}
    p_{\textrm{accept}} = \exp(-\chi^2(\theta_{n+1})/2 + \chi^2(\theta_n)/2) \,.
\end{equation}
This construction is known to have very large autocorrelations between the successive sets of accepted parameters. Also, the effectiveness of the sampling can be poor when the number of parameters in $\theta$ is large. In order to improve the performance, we use the estimated Hessian matrix to deduce correlations between the parameters. We note that this is only needed to decrease the computational cost and does not induce any bias in the sampling procedure, as each proposal is accepted or rejected according to the change of $\chi^2$. As described in the previous subsection, one can obtain the covariance matrix between the parameters from the Hessian matrix, i.e.
\begin{equation}
    \textrm{cov}_{ij} = \frac{1}{N} \sum_{k=1}^N \big( \theta^k_i - \bar{\theta}_i \big) \big( \theta^k_j - \bar{\theta}_j\big)\,, \qquad \textrm{where} \qquad  \bar{\theta_i} = \frac{1}{N} \sum_{k=1}^N \theta^k_i\,.
\end{equation}
Using the Cholesky decomposition, we define the matrix $\Omega$, such that
\begin{equation}
    \Omega^T \Omega = \textrm{cov}\,.
\end{equation}
This allows us to improve the proposal step of the MC algorithm (often called in this situation the adaptive MC algorithm), where the proposal is generated simultaneously for all the parameters from a Gaussian distribution,
\begin{equation}
    \theta_i \sim \mathcal{N}(0,1), \textrm{for each } i\,, \quad \textrm{and then} \quad  \tilde{\theta}_i = (\Omega^T)_{ij} \theta_j \quad \textrm{so that} \quad 
    \langle \tilde{\theta}_i \tilde{\theta}_j \rangle = \textrm{cov}_{ij}\,.
\end{equation}
The proposal $\theta$ is then accepted or rejected. The covariance matrix $\textrm{cov}$ can be further updated during the simulation. We observe that the acceptance rate, and hence the performance of the simulation, drastically increases when this improvement is included.

\subsection{Validation of the Bayesian MC approach}

\begin{figure*}[t] 
	\centering
    \includegraphics[width=0.5\textwidth]{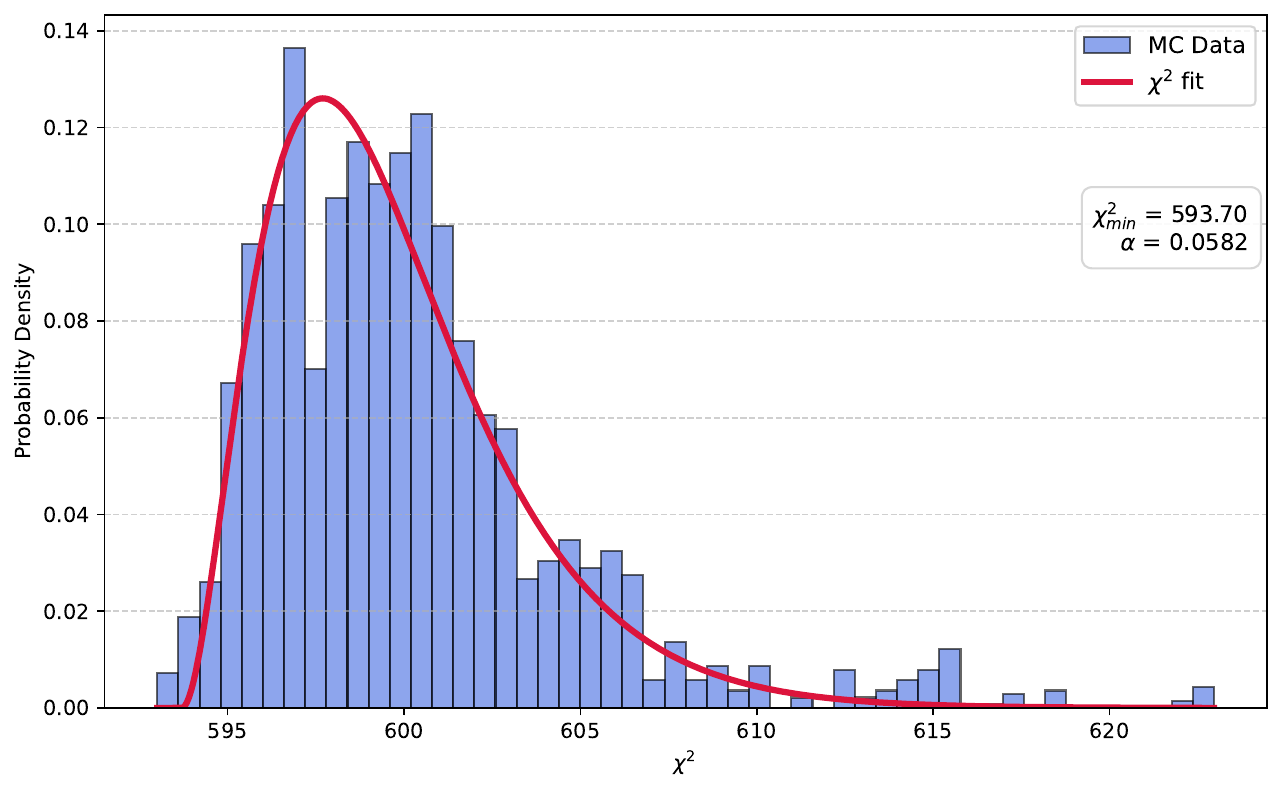}
    \caption{Validation of the Monte Carlo Markov chain. The histogram of the accepted $\chi^2$ values is plotted together with the theoretical expectation for this distribution with 6 parameters. The theoretical curve was fitted with two unknown parameters: $\chi^2_{\textrm{min}}$ and the normalization. We see good agreement confirming the correctness of the collected MC samples. }
    \label{fig:mcmc_histogram}
\end{figure*}

We implemented the Monte Carlo Bayesian inference approach for the simultaneous fit with 6 parameters. After rejecting the thermalization phase of the Markov chain, we collected 20000 samples that allowed us to extract the histograms for each parameter and cross-correlations discussed in the next subsection. As a test of the correctness of our results, we plot the histogram of accepted values of $\chi^2$. It is expected on theoretical grounds that the distribution of such values is given by the $\chi^2(n=6)$ distribution with 6 parameters. In principle, there is no free parameter left in this distribution. However, to cross-check the minimization results obtained by the optimization procedure, we leave the minimal value of $\chi^2$ function as a fit parameter, which we extract from the histogram. The results are plotted in Fig.~\ref{fig:mcmc_histogram}. We see that the extracted value of the minimal value of $\chi^2$ correctly corresponds to the value obtained by minimization. We also notice that the distribution approximated by the histogram corresponds nicely to the expected theoretical curve. 

\subsection{Comparison of the Bayesian MC and the Hessian methods}
\label{sec:MC_vs_hessian}

\begin{figure*}[t] 
	\centering
    \includegraphics[width=\textwidth]{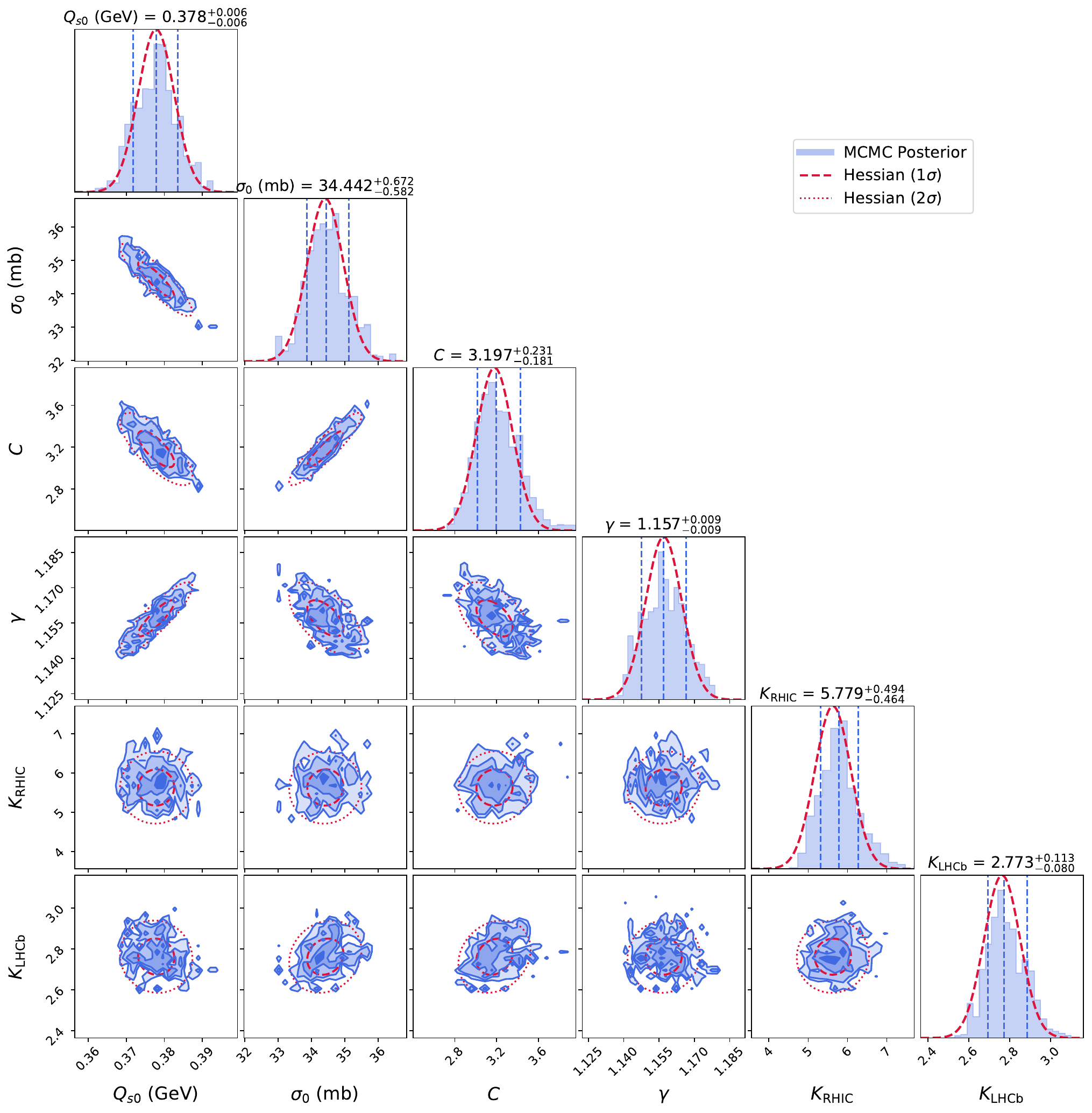}
    \caption{The comparison of the MCMC and Hessian methods. The result is calculated based on the global best fit. We see very good agreement between the distributions estimated by these two completely independent methods.}
    \label{fig:mcmc_hessian}
\end{figure*}

In Fig.~\ref{fig:mcmc_hessian}, we plot together the distributions for the fitted parameters and their correlations obtained using both the Hessian matrix and MC Bayesian inference methods. Note that these two methods are independent and do not involve additional tunable parameters. The comparison shows that both approaches yield very compatible results. This demonstrates that the parameters follow Gaussian distributions and therefore the determination relying only on the Hessian matrix is adequate. 

\end{document}